\documentclass[12pt]{article}
\usepackage{amssymb,amsmath,epsfig,cite}
 \allowdisplaybreaks

\begin{document}
\title{\bf Role of Structure Scalars on the evolution of Compact Objects in Palatini $f(R)$ Gravity}
\author{M. Z. Bhatti \thanks{mzaeem.math@pu.edu.pk}, Z. Yousaf
\thanks{zeeshan.math@pu.edu.pk}, and Z. Tariq \thanks{zohatariq24@yahoo.com}\\
Department of Mathematics, University of the Punjab,\\
Quaid-i-Azam Campus, Lahore-54590, Pakistan.}
\date{}
\maketitle
\begin{abstract}
The utmost concern of this article is the construction of modified scalar functions (structure scalars) by taking Palatini $f(R)$ gravitational theory into account. At first, a general formalism is established in which we assess gravitational stellar equations by putting into use the Palatini's technique. Later, from the perspective of tilted observer, we Lorentz boosted the components of energy-momentum tensor using relative velocity $\omega$. To examine the physical as well as mathematical aspects of the fluid source, we carry out a detailed analysis of kinematical variables by evaluating shear tensor and scalar, four-acceleration and expansion scalar. For the fluid content inside our spherical star, we inferred the mass function (geometric mass) and the active gravitational mass. Raychaudhuri equation, Bianchi identities in addition to few other equations are worked out to discern the structure formation and analyze the object's evolutionary stages. The Riemann tensor is then broken up orthogonally to set up few scalar functions connected with fundamental physical characteristics of the fluid source like energy density, effects of tidal forces and anisotropic stresses etc.
\end{abstract}

\section{Introduction}

Pursuant to the big bang theory, our universe was once an infinitely dense and extraordinarily hot small point resembling a mathematical infinity. Since, the mathematical infinities are irrational when it comes to physical equations, so the comprehension of universe at that infinite scale is laborious. Due to some unidentified reasons, the space started inflating, i.e., it endured an exponential expansion phase driving apart all the constituents of the universe. Owing to such expansion of the space, the temperature dropped down which resulted in the collision of protons and neutrons present in the space. This phenomenon produced Deuterium (an isotope of hydrogen gas) which in turn combined with each other to produce Helium (elementary ingredient for Star formation). Observations from NASA demonstrated that minor fluctuations in the matter density generated a gravitational pull that engendered an enormous cosmic web-like configuration of stars and space. Pulling in more and more material from space via this gravitational pull, the dense regions became more massive leading to the formation of stellar structures e.g., stars, galaxies, clusters and super-clusters of galaxies. Low density regions, on the other hand, showed no growth rather they evolved into emptiness known as void in space. The gravitational pull boosting the structure formation process was so intense that it decelerated the expansion phenomenon. Couple of years later, it was observed that universe is actually accelerating inexplicably, the root cause of which is an unidentified driving force termed as dark energy (DE). Its only physical characteristic identified till now is that it is a repulsive force exerting negative pressure on the objects.

Another oddity that prevails the space is the 'missing mass' required to maintain the stars orbiting around the galaxy's center in addition to the baryonic mass. Such mass (termed as dark matter (DM)) shows no interaction with electromagnetic radiation, light or any other detective source which makes it complicated to detect.
General relativity (GR) forecasted DE and DM but it cannot explicate both of them physically and mathematically. Many theoretical frameworks have been constructed \cite{36, 38} to account for them but no theory till now is able to entirely explicate them.  One of the credible approach is the introduction of substitutive gravitational theories hinged on the amendment of Einstein-Hilbert action integral by substituting a generic function $f(R)$ of Ricci scalar instead of Ricci scalar itself. Performing the variation of such action with the connection symbol and the metric tensor assuming their independence relation imparts the modified gravitational field equations which potentially explicate DM. Such theories, based upon the generic model employed, elucidates distinct eras of evolutionary phases of our universe.

Seifert \cite{1} investigated the stability of static solutions of a spherically symmetric matter content by making use of generalized variational principle in three substitutive gravitational theories. They concluded that $f(R)$ gravitational theory behaves like a highly unstable theory when the matter content is present.  Lobo \cite{2} reviewed certain alternative gravitational theories that cope with two current cosmological problems, DM and DE. They proposed that the galactic kinematics can be explicated using alternative gravitational theories without the introduction of DM terms. Multam\"{a}ki and Vijla \cite{3} studied the empty space solutions for static spherically symmetric matter content by taking into consideration the $f(R)$ theories of gravity. By reducing the number of modified field equations to one, they demonstrated that exact solutions can be rendered corresponding to distinct $f(R)$ gravity models. Yousaf et al. \cite{4} analyzed the dynamics of compact objects in modified theory via structure scalars. Taking into consideration a heat emitting shearing matter content, they evaluated structure scalars by breaking up the Riemann tensor orthogonally in $f(R,T)$ gravity. It is inferred that by making use of the modified scalar functions, the evolutionary behavior of realistic astrophysical objects can be discussed. Olmo \cite{5} over-viewed the Newtonian framework and devoted their attention towards Nordstr\"{o}m's scalar theory and its impact on Einstein's GR. They also presented the cosmic speed-up issue and inferred that several quantum gravity issues can be well addressed by the Palatini $f(R)$ gravity and $f(R,Q)$ theories \cite{5q}.

In order to establish consistent and viable models of relativistic bodies, the interior region is matched with the exterior region at a boundary known as hypersurface. Such matching leads to certain conditions termed as junction conditions. Olmo and Garcia \cite{6} evaluated the matching conditions in $f(R)$ gravity using a tensor distributional technique and inferred that few of such restrictions deviate from standard ones. They also illustrated the significance of these restrictions by taking into consideration various characteristics of stellar objects in polytropes. Their work can be used to safely model Neutron stars and White dwarfs in Palatini $f(R)$ gravity. Yousaf et al. described the role of modified gravity terms including Palatini $f(R)$ \cite{6a,6b,6c} and $f(G,T)$ \cite{6d, 6e,6f} corrections on the stability of homogeneous and inhomogeneous relativistic stars. Senovilla \cite{7} inferred matching conditions in modified gravity i.e. $f(R)$ to explicate matter shells and braneworld. For this purpose, author acquired the modified field equations for the stress-energy tensor on the shell and deduced that it was similar to that found for GR. Huber \cite{8} proposed a framework based on the geometric deformations to address the issue of joining two spacetime manifolds having distinct physical characteristics and geometries. The author exhibited that such a geometric technique assists in the depiction of spacetimes in GR. Deruelle et al. \cite{9} generalized the matching conditions for $f(R)$ theories by directly integrating the Einstein's equations. This is possible because of the equivalent relation between GR coupled to scalar field and $f(R)$ theory. Bonnor and Vickers \cite{10} analyzed certain junction conditions in GR and showed the existence of equivalence relation between few of them. Out of all the conditions, it was discovered that the Darmois conditions are the most trustworthy and easy to handle.

An algebraic expression that assists in assessment of total energy of the gravitating body is called the active gravitational mass. Herrera et al. \cite{11} acquired an algebraic expression for the Tolman mass of a collapsing matter content. This expression brings into light the part played by the energy density inhomogeneity and anisotropy in the collapsing process of the spherical object. It was deduced that the energy density inhomogeneity increases in the Tolman mass expression during the collapsing phenomena and local anisotropy assists in the development of singularities. Bonnor \cite{12} calculated the active and passive gravitational masses of a static spherically symmetric fluid content having uniform density. The results exhibit that the active gravitational mass is much smaller than the passive one. The dynamics of such a sphere are discussed in detail. Devitt and Florides \cite{13} worked out an expression for modified active gravitational mass of a spherically symmetric fluid content having no dependence on temporal coordinate. They enlisted certain characteristics of the modified formula thus obtained including the fact that such an expression provides the correct mass of the overall system. Herrera et al. \cite{14} attained an algebraic expression for the Tolman mass of a source just after its withdrawal   from equilibrium phase. For the case of compact sources, it is noticed that small withdrawal from sphericity results in a substantial change in Tolman mass of the considered system as compared to the value in equilibrium phase. Misner and Putnam \cite{15} scrutinized a paradox based on the active gravitational mass of a spherical source for the case when interior gravitational influences are insignificant.

The changes induced in a star's interior with the passage of time is termed as stellar evolution process. Schaab et al \cite{16} applied the theoretical equation of state to analyze the cooling characteristics of compact astronomical objects and tested a number of uncertainties including the dynamics of coupling strength of matter and the super fluidity.  Althaus et al. \cite{17} reviewed the physical characteristics of white dwarfs with focus on the processes involved in their formation. The pulsational characteristics of white dwarfs and their implications in modern cosmology are also investigated. Chabrier and Baraffe \cite{18} analyzed the structural formation and evolutionary behavior of low-mass stars with special focus on the thermal and mechanical characteristics of such objects. They also demonstrated mass-radius luminosity relationship via numerical tabulation for different ages of these stars and observed that the minimum masses are much smaller than previous approximations. Sakashita and Hayashi \cite {19} computed few models of highly massive starts and concluded that such stars evolve in a same manner as low-mass stars after their formation.

A persistent battle between the force of gravity and pressure helps a stellar object in maintaining its physical state. When one of these forces supersedes the other one, stellar collapse occurs. Astashenok et al. \cite{20} carried out a comparative analysis of the collapsing phenomena for perfect homogeneous matter content via distinct equations of state in GR and $R$-squared gravitational theory. Considering the cases of radiating and stiff matter, they also found the possibility of presence of vacuum energy along with its impact on the collapsing phenomena. Chiba \cite{21} explored the dust collapse for a cylindrical symmetry and presented a novel theoretical approach to analyze the singularity formation in such symmetry. They concluded that negligible gravitational waves are emitted within the free fall time. Taking into account the quantum mechanical influences, Balakrishna et al \cite{22} modeled a collapsing star into a black hole. They derived certain solutions for the path of the practical on the collapsing star's surface in crucial and Schwarzschild coordinators. Fryer and New \cite{23} over-viewed all the possible collapsing sources of gravitational waves including the collapsing phenomena of white dwarf. They also investigated the collapse of Black holes and neutron stars in addition with other super massive stars.

For a detailed examination of fluid characteristics, few kinematical variables are ascertained. These include the four-acceleration, the vorticity vector, shear tensor and expansion scalar etc. Keeping in view the quasistatic approximation, Herrera and Santos \cite{24} inquired the shear-free restriction for the case of dissipative self-gravitating matter content. They inferred that the expansion rate in addition to the shear-free restriction are equivalent to the homology restriction. They also worked out the implications of such models to some astronomical scenarios. Yousaf et al. \cite{25} examined the irregularity factors for a spherically symmetric star by taking into consideration an imperfect fluid content. They further investigated two differential equations utilizing the conservation laws and the Weyl tensor and concluded that the complexity of object increases with the increase in anisotropic stresses. Herrera et al. \cite{26} revisited the notion of titled congruence for Lemaitre-Tolman bondi spacetime with an imperfect fluid distribution. They evaluated the factors responsible for the presence of energy density inhomogeneities and carried out a comparison with the outcomes obtained for the case of non tilted congruence. Herrera et al. \cite{27} demonstrated the instability of spherical systems undergoing non-adiabatic   collapse with radial heat dissipation. They presented that the Newtonian correction causes of heat flow maximizes the fluid instability whereas the relativistic correction lessens the instability of the fluid. Herrera et al. \cite{28} explored on the expansion-free spherically symmetric stars by evaluating the field equations along with the junction conditions, for an anisotropic fluid with heat flux to comprehend the expansion-free motion. They defined radial velocity of matter in two distinct ways and analyzed each in detail.

Few of the factors that describe the physical state of fluid source create complications in the analysis of gravitating relativistic bodies. These factors are called complexity factors. Herrera \cite{29} proposed a novel concept of complexity for the case of time independent spherically symmetric objects. They determined certain exact solutions to the Einstein's stellar equations fulfilling the zero complexity factor criterion and also found implications of this novel concept of the analysis of structure development of compact object. Herrera et al. \cite{30} propounded a systematic study of complexity for time independent system with spherical symmetry along with considering the complexity factor of fluid content. They also obtained the minimal complexity condition analyzing the dissipative and non-dissipative cases in detail and also explored the stability of zero complexity condition. Yousaf \cite{31} revisited the complexity factor for relativistic spheres having non-dynamical equation in Palatini $f(R)$ gravitational theory. Manipulation of the field equations along with the TOV equation assists in constructing the complexity factor of the system. The part played by the $f(R)$ terms and fluid variables is also analyzed in the discussion of evolution of such systems. Sa\~{n}udo and Pacheo \cite{32} studied the density profiles of white dwarfs in order to evaluate its complexity. They inferred that the complexity of such a structure increased with the value of atomic number. Chatziavvas et al. \cite{33} applied the notion of measuring complexity statistically to a Neutron  star model in continuation to the work of \cite{32} from the analysis of gravity and short range force. They observed that under their theoretical framework. The Neutron stars less complex objects. Bhatti and Tariq \cite{34,35} carried out a study of spherically symmetric astronomical object with anisotropy in fluid having heat dissipation in the presence and absence of electromagnetic field in both GR and $f(R)$ modified gravitational theories. They manipulated certain evolution equation to demonstrate the physical processes involved in the evolutionary phases of such systems via structure scalars.

The structure of this research article is organized in the following pattern: Section \textbf{1} encompasses Palatini $f(R)$ field equations, the dynamical quantities expressing fluid properties and the Junction conditions. Misner-Sharp mass, electric part of the Weyl tensor and the active gravitational mass are evaluated in Section \textbf{2}. In order to comprehend how the fluid evolves with time, certain evolution equations are also worked out. The Riemann tensor is divided into three tensor quantities $X_{\varrho\beta}, Y_{\varrho\beta}$ and $Z_{\varrho\beta}$ in Section \textbf{3} to acquire few scalars connected with the fluid dynamics. Such scalars are then interlinked with few equations of motion to get an idea about how the fluid evolves with time. Section \textbf{4} enlists three locally anisotropic static spheres. Section \textbf{5} incorporates outlines of the acquired results.

\section{Palatini Field Equations along with Kinematical Variables}

Amendment in the gravitational constituent of the Einstein-Hilbert action renders the $f(R)$ gravitational theory as follows
\begin{equation}\label{1d}
S_{f(R)}=\frac{1}{2\kappa}\int f(R)\sqrt{-g}d^4x+S_M,
\end{equation}
where $f(R)$ represents a non-linear generic Ricci function and $\kappa$ and $S_M$ are referred to as the coupling constant and the matter action respectively. In order to examine a dissipative fluid content having spherical symmetry and enduring the collapsing phenomenon under anisotropic stresses, we specify the following metric
\begin{equation}\label{2d}
ds^2=e^{\nu(t,r)} dt^2- e^{\lambda(t,r)} dr^2-r^2(d\theta^2+sin^2 \theta d\phi^2).
\end{equation}
Exploiting the Palatini's technique, the following composition of Palatini gravitational equations is attained
\begin{align}\nonumber
&\frac{1}{f_R}\left(\nabla_\mu \nabla_\delta f_R-g_{\mu\delta}\Box f_R\right)+\frac{\kappa T_{\mu\delta}}{f_R}-\frac{1}{2}g_{\mu\delta}\left(R-\frac{f}{f_R}\right)+\frac{3}{2}f_R^2\left(\frac{1}{2}g_{\mu\delta}(\nabla f_R)^2 -\nabla_\mu f_R \nabla_\delta f_R\right)\\\nonumber &=R_{\mu\delta}-\frac{1}{2}g_{\mu\delta}R.
\end{align}
Compiling this equation in an Einstein-like structure of field equations, we infer
\begin{equation}\label{3d}
G^\mu_{\delta}=\frac{\kappa}{f_R}\left[T^\mu_{\delta}+\mathcal{T}^\mu_{\delta}\right],
\end{equation}
where $\mathcal{T}_{\mu\delta}$ symbolizes the effective energy-momentum tensor composed of pure geometric elements and is provided below
\begin{equation}\nonumber
\mathcal{T}^\mu_{\delta}=\frac{1}{\kappa}\left(\nabla^\mu \nabla_\delta-\delta^\mu_{\delta}\Box \right)f_R+\frac{\delta^\mu_{\delta} f_R}{2\kappa}+\frac{3}{2\kappa f_R}\left[\frac{1}{2}\delta^\mu_{\delta}(\nabla f_R)^2 -\nabla^\mu f_R \nabla_\delta f_R\right]
\left(\frac{f}{f_R}-R\right).
\end{equation}
Equation (\ref{3d}) leads us to the following Palatini gravitational stellar equations
\begin{align}\label{4d}
\kappa T^0_0&=f_R\left[\frac{1}{r^2}-\frac{e^-\lambda}{r^2}+\frac{\lambda'e^{-\lambda}}{r}\right]+\bar{\tau}_0,\\\label{5d}
\kappa T^1_1&= f_R\left[\frac{1}{r^2}-\frac{e^{-\lambda}}{r^2}-\frac{\nu' e^{-\lambda}}{r}\right]+\bar{\tau_1},\\\nonumber
\kappa T^2_2&= f_R\left[\frac{\ddot{\lambda} e^{-\nu}}{2}+\frac{\dot{\lambda^2}e^{-\nu}}{4}
-\frac{\dot{\lambda }\dot{\nu} e^{-\nu}}{4}+\frac{\lambda' e^{-\lambda}}{2r}-\frac{\nu'' e^{-\lambda}}{2}-\frac{\nu^{'2}e^{-\lambda}}{4}-\frac{\nu' e^{-\lambda}}{2r}+\frac{\lambda' \nu' e^{-\lambda}}{4}\right]+\bar{\tau_2},\\\label{6d} \kappa T_{10} &= \frac{\dot{\lambda}f_R}{r}+\bar{\tau_3}.
\end{align}
The dark source terms arising due to Palatini $f(R)$ emendations are
\begin{align}\nonumber
\bar{\tau}_0&=\frac{9e^{-\nu}\dot{f_R}^2}{4f_R}+\frac{e^{-\lambda}f_R'^2}{4f_R}-e^{-\lambda}f_R''+
\frac{\dot{\lambda}e^{-\nu}\dot{f_R}}{2}-\frac{\lambda'e^{-\lambda}f_R'}{2}-\frac{2e^{-\lambda f_R'}}{r}+\frac{f_R}{2}\left(R-\frac{f}{f_R}\right),\\\nonumber
\bar{\tau}_1&=\frac{-e^{-\nu}\dot{f_R}^2}{4f_R}-\frac{9 e^{-\lambda} f_R'^2}{4f_R}+\ddot{f_R}e^{-\nu}-\frac{\dot{\nu}\dot{f_R}e^{-\nu}}{2}-\frac{\nu' e^{-\lambda} f_R'}{2}-\frac{2e^{-\lambda} f_R'}{r}-\frac{f_R}{2}(R-\frac{f}{R}),\\\nonumber
\bar{\tau}_2&=-\frac{e^{-\nu} \dot{f_R}^2}{4f_R}+\frac{e^{-\nu}\dot{f_R}^2}{4f_R}-\frac{e^{-\lambda}f_R'}{r}+e^{-\nu}\ddot{f_R}-\frac{\dot{\nu}\dot{f_R} e^{-\nu}}{2}-\frac{\nu' e^{-\lambda} f_R'}{2}-e^{-\lambda} f_R''+\frac{\dot{\lambda} e^{-\lambda}\dot{f_R}}{2}\\\nonumber &+
\frac{\lambda' e^{-\lambda}f_R'}{2}+\frac{f_R}{2}(R-\frac{f}{f_R}),\\\nonumber
\bar{\tau}_3& = -\dot{f_R'}+\frac{\nu' \dot{f_R}}{2}+\frac{\dot{\lambda}f_R'}{2}+\frac{5\dot{f_R}f_R'}{2f_R}.
\end{align}
The dotted and primes values of the metric coefficients signify the temporal and radial differentiation respectively. \\
The gravitational influences from the fluid sources decides the geometry of the astronomical bodies. Four-velocities of such fluid sources are the elementary constituents in the composition of energy-momentum tensor. If two distinct relativistic explications of a spacetime manifold are interconnected with the boost of one observer relative to the other, then the kinematics of the fluid sources can be different. Keeping in view the work of Bondi \cite{39}, we write the Minkowskian coordinates as
\begin{eqnarray}\nonumber
d\tilde{\eta}=e^{\nu/2}dt; \quad d\tilde{y}=rd\theta;\quad
d\tilde{x}=e^{\lambda/2}dr; \quad d\tilde{z}=r sin\theta d\phi.
\end{eqnarray}
To exploit the tilted congruence notion, it is supposed that the fluid source has radial velocity $\omega$ in accordance with the new reference frame. We employ Lorentz boost from Minkowskian frame to new frame. So, we write Minkowskian energy-momentum tensor components as
\begin{eqnarray}\nonumber
\check{T}_0^0=T_0^0; \quad\check{T}_1^1=T_1^1;\quad
\check{T}_2^2=T_2^2; \quad \check{T}_3^3=T_3^3;\quad
\check{T}_{01}=e^{-(\nu+\lambda)/2}T_{01} .
\end{eqnarray}
Thus, a comoving observer observes the following covariant energy-momentum tensor components
$$
\begin{pmatrix}
\rho+\varepsilon & -q-\varepsilon & 0 & 0\\
-q-\varepsilon & P_r+\varepsilon & 0 & 0\\
0 & 0 & P_\bot & 0\\
0 & 0 & 0 & P_\bot \\
\end{pmatrix},
$$
with $\varepsilon, \rho, q, P_r$ and $P_\bot$ signify radiation and energy density, the heat flux vector and the radial and tangential
stresses respectively. Putting the idea of Lorentz transformation in use, we render
\begin{align}\label{7d}
\check{T}_0^0&=T_0^0=\frac{\rho+P_r
\omega^2}{1-\omega^2}+\frac{2\omega
q}{1-\omega^2}+\frac{\varepsilon(1+\omega)}{1-\omega},\\\label{8d}
\check{T}_1^1&=T_1^1=-\left(\frac{\rho
\omega^2+P_r}{1-\omega^2}+\frac{2\omega
q}{1-\omega^2}+\frac{\varepsilon(1+\omega)}{1-\omega}\right),\\\nonumber
\check{T}_{01}=e^{-(\nu+\lambda)/2}T_{01}&=-\left(\frac{\omega
e^{(\nu+\lambda)/2}(\rho+P_r)}{1-\omega^2}+\frac{qe^{(\nu+\lambda)/2}(1+\omega
^2)}{1-\omega^2}\right.\\\label{9d}
&\left.+\frac{\varepsilon(1+\omega)e^{(\nu+\lambda)/2}}{1-\omega}\right),\\\label{10d}
\check{T}_2^2&=T_2^2=-P_\bot; \quad \check{T}_3^3=T_3^3=-P_\bot.
\end{align}
For the fluid source velocity $\omega$, we have
\begin{equation}\nonumber
\omega=\left(\frac{dr}{dt}\right)e^{\frac{(\lambda-\nu)}{2}}.
\end{equation}
Substituting Eqs. (\ref{7d}-\ref{10d}) into Eqs. (\ref{3d}-\ref{6d}), we attain
\begin{align}\label{11d}
&\frac{\rho+P_r \omega^2}{1-\omega^2}+\frac{2\omega
q}{1-\omega^2}+\frac{\varepsilon(1+\omega)}{1-\omega}=\frac{f_R}{\kappa}\left(\left[\frac{\lambda'
e^{-\lambda}}{r}+\frac{1}{r^2}-\frac{e^{-\lambda}}{r^2}\right]
+\frac{\bar{\tau}_0}{f_R}\right),\\\label{12d}
&\left(\frac{\rho \omega^2+P_r}{1-\omega^2}+\frac{2\omega
q}{1-\omega^2}+\frac{\varepsilon(1+\omega)}{1-\omega}\right)=-\frac{f_R}{\kappa}\left(\left[-\frac{\nu'
e^{-\lambda}}{r}+\frac{1}{r^2}-\frac{e^{-\lambda}}{r^2}\right]
+\frac{\bar{\tau}_1}{f_R}\right),\\\label{13d}
&P_\bot=-\frac{f_R}{\kappa}\left(\left[e^{-\nu}\left(\frac{\ddot{\lambda}}{2}+\frac{\dot{\lambda}^2}{4}-\frac{\dot{\lambda}
\dot{\nu}}{4}\right)+e^{-\lambda}\left(\frac{{\lambda}'}{2r}-\frac{{\nu}''}{2}-\frac{{\nu}'^2}{4}-\frac{{\nu}'}{2r}+\frac{{\lambda}'
{\nu}'}{4}\right)\right]+\frac{\bar{\tau}_2}{f_R}\right),\\\label{14d}
&\left(\frac{\omega
e^{(\nu+\lambda)/2}(\rho+P_r)}{1-\omega^2}+\frac{qe^{(\nu+\lambda)/2}(1+\omega
^2)}{1-\omega^2}+\frac{\varepsilon(1+\omega)e^{(\nu+\lambda)/2}}{1-\omega}\right)=
-\frac{f_R}{\kappa}\left(\frac{\dot{\lambda}}{r}+\frac{\bar{\tau}_3}{f_R}\right).
\end{align}
The notion of tilted congruence supports the use of following four-velocity vector
\begin{equation}\nonumber
u^\nu=\left(\frac{e^{-\nu/2}}{(1-\omega^2)^{1/2}},\frac{\omega e^{-\lambda/2}}{(1-\omega^2)^{1/2}},0,0\right)
\end{equation}

\textbf{Kinematical Variables}

Being already familiar with the expression for four-acceleration i.e. $a^\eta=u^\eta_{;\lambda} u^\lambda$, we evaluate its surviving components as
\begin{align}\nonumber
a_0&=\frac{\nu' \omega e^{\frac{\nu-\lambda}{2}}}
{2(1-\omega^2)}+\frac{\omega \dot{\omega}}{(1-\omega^2)^2}+\frac{\omega^2 \omega' e^{\frac{\nu-\lambda}{2}}
}{(1-\omega^2)^2}+\frac{\dot{\lambda} \omega^2}{2(1-\omega^2)}+\frac{w f_R' e^{\frac{\nu-\lambda}{2}}}{2f_R (1-\omega^2)}+\frac{\dot{f_R}(1+\omega^2)}{2f_R (1-\omega^2)},\\\nonumber
a_1&=-\frac{\nu'}{2(1-\omega^2)}-\frac{\dot{\omega}e^{\frac{\lambda-\nu}{2}}}{(1-\omega^2)^2}-\frac{\dot{\lambda}\omega e^{\frac{\lambda-\nu}{2}}}{2(1-\omega^2)}-\frac{\omega\omega'}{(1-\omega^2)^2}-\frac{f_R'(1+\omega^2)}{2f_R (1-\omega^2)}
-\frac{\dot{f_R} \omega e^{\frac{\lambda-\nu}{2}}}{f_R(1-\omega^2)}.
\end{align}
To explicate the physical characteristics of the fluid content, we manipulate the shear tensor components $(\sigma_{\beta\eta}=u_{\beta;\eta}+u_{\eta;\beta}-u_\beta
a_\eta-u_\eta a_\beta-\frac{2\Theta h_{\beta\eta}}{3})$ along with the evaluation of expansion scalar $(\Theta=u^\eta_{; \eta})$ as follows
\begin{align}\nonumber
\Theta&=\frac{\dot{\lambda}e^{-\nu/2}}{2\sqrt{(1-\omega^2)}}+
\frac{\omega\dot{\omega}e^{-\nu/2}}{(1-\omega^2)^{3/2}}+
\frac{\omega\nu'e^{-\lambda/2}}{2\sqrt{(1-\omega^2)}}+
\frac{\omega'e^{-\lambda/2}}{(1-\omega^2)^{3/2}}+ \frac{2\omega
e^{-\lambda/2}}{r\sqrt{(1-\omega^2)}}\\\nonumber &+\frac{2f_R' \omega e^{-\lambda/2}}{f_R (1-\omega^2)^{1/2}}+\frac{2\dot{f_R}e^{-\nu/2}}{f_R \sqrt{(1-\omega^2)}},
\end{align}
and
\begin{align}\nonumber
\sigma_{00}&=-\frac{4\omega^3\dot{\omega}e^{\nu/2}}{3(1-\omega^2)^{5/2}}-\frac{2\nu' \omega^3 e^{\nu-\lambda/2}}{3(1-\omega^2)^{3/2}}
-\frac{4\omega^2\omega'e^{\nu-\lambda/2}}{3(1-\omega^2)^{5/2}}-\frac{2\dot{\lambda}\omega^2 e^{\nu/2}}{3(1-\omega^2)^{3/2}}
+\frac{4\omega^3 e^{\nu-\lambda/2}}{3r(1-\omega^2)^{3/2}}\\\nonumber&-\frac{2\dot{f_R}e^{\nu/2}}{f_R(1-\omega^2)^{1/2}}
-\frac{2\omega^2\dot{f_R}e^{\nu/2}}{3f_R(1-\omega^2)^{3/2}}+\frac{4\omega^3f_R'}{3f_R(1-\omega^2)^{3/2}},\\\nonumber
\sigma_{11}&=-\frac{4\omega'e^{\lambda/2}}{3(1-\omega^2)^{5/2}}-\frac{2\dot{\lambda}e^{\frac{\lambda-\nu}{2}}}{3(1-\omega^2)^{3/2}}
+\frac{\dot{f_R}e^{\frac{\lambda-\nu}{2}}}{3 f_R(1-\omega^2)^{5/2}}-\frac{2\omega f_R'e^{\lambda/2}}{3 f_R(1-\omega^2)^{3/2}}-\frac{4\omega\dot{\omega}e^{\frac{\lambda-\nu}{2}}}{3(1-\omega^2)^{5/2}}\\\nonumber&-\frac{2\nu'\omega e^{\lambda/2}}{3(1-\omega^2)^{5/2}}+\frac{4\omega e^{\lambda/2}}{3r(1-\omega^2)^{3/2}},\\\nonumber
\sigma_{22}&=\frac{r^2e^{-\nu/2}\dot{f_R}}{3f_R(1-\omega^2)^{1/2}}+\frac{\omega r^2 e^{-\lambda/2}f_R'}{3f_R(1-\omega^2)^{1/2}}
-\frac{2\omega re^{-\lambda/2}}{3(1-\omega^2)^{1/2}}+\frac{2r^2 \omega\dot{\omega}e^{-\nu/2}}{3(1-\omega^2)^{3/2}}+\frac{2r^2 \omega'e^{-\lambda/2}}{3(1-\omega^2)^{3/2}}\\\nonumber &+\frac{r^2 \nu' \omega e^{-\lambda/2}}{3(1-\omega^2)^{1/2}}+\frac{\dot{\lambda}r^2 e^{-\nu/2}}{3(1-\omega^2)^{1/2}},\\\nonumber
\sigma_{33}&=\sigma_{22} Sin^2 \theta,\\\nonumber
\sigma_{01}&=\frac{2\omega \lambda e^\frac{\lambda-\nu}{2}}{3(1-\omega^2)^{3/2}}+\frac{4 \omega^2 \dot{\omega}e^\frac{\lambda-\nu}{2}}{3(1-\omega^2)^{5/2}}
+\frac{2\omega^2 \nu' e^{\nu/2}}{3(1-\omega^2)^{3/2}}+\frac{4\omega \omega' e^{\nu/2}}{3(1-\omega^2)^{5/2}}+\frac{4\omega e^{\lambda/2}}{3_r(1-\omega^2)^{3/2}}\\\nonumber &+\frac{7\omega^2 f_R' e^{\nu/2}}{6f_R(1-\omega^2)^{3/2}}+\frac{7\dot{f_R}\omega e^{\lambda/2}}{6f_R(1-\omega^2)^{3/2}}-\frac{\dot{f_R}\omega^3 e^{\lambda/2}}{2f_R(1-\omega^2)^{3/2}},
\end{align}
where $h_{\beta\eta}=g_{\beta\eta}-u_\beta u_\eta$. The shear scalar using the Palatini's technique renders the following outcome
\begin{align}\nonumber
\sigma&=\frac{2\omega e^{-\lambda/2}}{r(1-\omega^2)^{1/2}}-\frac{\dot{\lambda}e^{-\nu/2}}{(1-\omega^2)^{1/2}}\frac{-2\omega' e^{\lambda/2}}{(1-\omega^2)^{3/2}}-\frac{3\dot{f_R}e^{-\nu/2} (1-\omega^2)^{1/2}}{f_R \omega^2}+\frac{2\omega f_R' e^{-\lambda/2}}{f_R(1-\omega^2)^{1/2}}\\\nonumber &-\frac{2\omega \dot{\omega}e^{-\nu/2}}{(1-\omega^2)^{3/2}}-\frac{\nu' \omega e^{-\lambda/2}}{(1-\omega^2)^{1/2}}.
\end{align}
Another four-vector field with the following definition is presumed
\begin{equation}\label{15d}
s^\nu=\left(\frac{\omega e^{-(\nu/2)}}{(1-\omega^2)^{1/2}}, \frac{e^{-\lambda/2}}{(1-\omega^2)^{1/2}},0,0\right).
\end{equation}
Pressure anisotropy has crucial contribution in the study of collapsing astronomical objects as it increases the complexity in them. For our work, we choose anisotropic fluid source that can be indicated by the following energy-momentum tensor
\begin{equation}\nonumber
T^\beta_\eta=\tilde{\rho}u^\beta u_\eta -\check{P}h^\beta_\eta
+\Pi^\beta_\eta + \tilde{q}(s^\beta u_\eta+s_\eta u^\beta),
\end{equation}
with
\begin{align}\nonumber
\Pi^\beta_\eta&=\Pi\left(s^\beta
s_\eta+\frac{h^\beta_\eta}{3} \right); \quad
\check{P}=\frac{2P_\bot +\tilde{P}_r}{3};\quad
\tilde{q}^\gamma=\tilde{q}s^\gamma; \quad
\tilde{\rho}=\rho+\varepsilon; \\\nonumber
\tilde{P_r}&=P_r+\varepsilon;\quad \tilde{q}=q+\varepsilon;
\quad\Pi=\tilde{P}_r-P_\bot.
\end{align}
In order to discern the singularity areas of curvature and matter contents at hypersurface $\Sigma$, we exploit the junction conditions and determine the permitted values for discontinuity of certain quantities across $\Sigma$. The radiating Schwarzschild metric i.e. the Vaidya's metric given below is considered for this purpose
\begin{equation}\nonumber
ds^2=\left(1-\frac{2M(\bar{\eta})}{\Re}\right) d\bar{\eta}^2 +2d\bar{\eta} d\Re -\Re^2 (d\theta^2+sin^2\theta d\phi^2).
\end{equation}
For our spherically symmetric object, the Darmois junction conditions are determined as
\begin{align}\nonumber
e^{\nu_\Sigma}=\left(1-\frac{2M(\bar{\eta})}{\Re_\Sigma}\right),
e^{-\lambda_\Sigma}=\left(1-\frac{2M(\bar{\eta})}{\Re_\Sigma}\right),\quad[(P_r)_{eff}=(q)_{eff}]_\Sigma.
\end{align}

\subsection{The Curvature Tensors (Weyl and Riemann tensors)}

It is a formerly known fact that the fluid content engenders a distortion (curvature) in spacetime. We can measure the extent of such curvature with the help of curvature tensors i.e the Riemann and the Weyl tensor. For a spherical star, the Weyl tensor comprises only the electric part as the magnetic part becomes zero due to symmetric property. From the expression used to explicate the Weyl tensor, we can extract the definition of Riemann tensor too. This is provided as follows
\begin{equation}\label{16d}
R^\nu_{\eta\rho\mu}=C^\nu_{\eta\rho\mu}+\frac{1}{2}R^\nu_\rho
g_{\eta\mu}+\frac{1}{2}R_{\eta\rho}\delta^\nu_\mu+\frac{1}{2}R_{\eta\mu}\delta^\nu_\rho-\frac{1}{2}R^\nu_\mu
g_{\eta\rho}-\frac{1}{6}R\left(\delta^\nu_\rho
g_{\eta\mu}-g_{\eta\rho}\delta^\nu_\mu\right).
\end{equation}
The Weyl tensor in terms of its electric part is defined as
\begin{equation}\nonumber
C_{\xi\nu\pi\lambda}=E^{\beta\delta}u^\rho u^\gamma
(g_{\xi\nu\rho\beta}
g_{\pi\lambda\gamma\delta}-\eta_{\xi\nu\rho\beta}
\eta_{\pi\lambda\gamma\delta}),
\end{equation}
where $g_{\xi\nu\rho\beta}$ is equal to
$g_{\xi\nu\alpha\beta}=g_{\xi\rho}g_{\nu\beta}-g_{\xi\beta}g_{\nu\rho}$
and $\eta_{\pi\lambda\gamma\delta}$ indicates Levi-Civita
tensor. $E_{\varrho\beta}$ has the following alternative form
\begin{equation}\label{17d}
E_{\varrho\beta}=E\left(s_\varrho s_\beta+\frac{1}{3} h_{\varrho\beta}
\right).
\end{equation}
The Weyl scalar is determined to be as follows
\begin{align}\nonumber
E&=\frac{\ddot{\lambda}e^{-\nu}}{4}+\frac{\dot{\lambda}^2
e^{-\nu}}{8}-\frac{\dot{\lambda}\dot{\nu}
e^{-\nu}}{8}-\frac{{\nu}''e^{-\lambda}}{4}-\frac{\nu'^2
e^{-\lambda}}{8}+\frac{\lambda'\nu'e^{-\lambda}}{8}+\frac{\nu'e^{-\lambda}}{4r}-
\frac{\lambda'e^{-\lambda}}{4r}-\frac{e^{-\lambda}}{2r^2}+\frac{1}{2r^2},
\end{align}
with $E$ fulfilling the following criteria
\begin{equation}\nonumber
E^\eta_\eta=0; \quad\quad E_{\eta\sigma}=E_{(\eta\sigma)}; \quad\quad
E_{\eta\sigma}u^\sigma=0.
\end{equation}

\subsection{Mass Function for the Spherical Interior}

The interior geometric mass is inferred as
\begin{equation}\label{18d}
R^3_{232}=-e^{-\lambda}+1=\frac{2m}{r}.
\end{equation}
Putting into use the Weyl scalar, the field equations attained from Platini's variational technique and the definition of Reimann tensor provided in Eq. (\ref{16d}),we acquire
\begin{equation}\label{19d}
\frac{3m}{r^3}=E+\frac{\kappa}{2f_R} (\tilde{\rho}-\tilde{P}_r +P_\bot)-\frac{1}{2f_R}(\bar{\tau}_0+\bar{\tau_1}).
\end{equation}
Eqs. (\ref{4d}) and (\ref{18d}) lead us to the following expression for the interior geometric mass
\begin{equation}\label{20d}
m=\frac{1}{2}\int^r_0\frac{\kappa r^2}{f_R}\left(T^0_0-\frac{\bar{\tau}_0}{\kappa}\right).
\end{equation}
This result when compared with Eq. (43) in \cite{41} demonstrates that the appearance of the dark source term $\bar{\tau}_0$ emerging due to Palatini's technique affects the interior mass of the astronomical body. Using Eqs. (\ref{16d}) and (\ref{17d}) in addition with Eq. (\ref{20d}) and field equations, we infer
\begin{equation}\label{21d}
m=\frac{ r^3}{6f_R}\kappa(T^1_1 +T^0_0 -T^2_2)+\frac{E r^3 }{3}-\frac{(\bar{\tau_0}+\bar{\tau_1})r^3}{6f_R}.
\end{equation}
The radial differentiation of this equation in combination with Eq. (\ref{20d}) provides
\begin{align}\label{22d}
\left(\frac{r^3}{3} E\right)'&=-\frac{ r^3}{6f_R}\left[\kappa(T^0_0)'+\frac{3\bar{\tau}_0}{ r}\right]+\left[\frac{ r^3}{6f_R}\kappa (T^2_2-T^1_1)\right]'+\left[\frac{(\bar{\tau}_0+\bar{\tau}_1)r^3}{6f_R}\right]'.
\end{align}
Carrying out the integration of above equation, we acquire
\begin{align}\label{23d}
E&=-\frac{1}{2r^3}\int^r_0\frac{ r^3}{f_R}\left[\kappa(T^0_0)'+\frac{3\bar{\tau}_0}{ r}\right]dr+\frac{1}{2f_R}\kappa(T^2_2-T^1_1)+\frac{\bar{\tau}_0+\bar{\tau}_1}{2f_R}.
\end{align}
This outcome has correspondence with Eq. (46) in \cite{41} with a distinction that the tidal force for our theory, in addition to the energy density inhomogeneity and stress anisotropy, is also dependent on the dark source quantities $\bar{\tau}_0$ and $\bar{\tau}_1$.
Placing the value of $E$ from Eq. (\ref{23d}) into  Eq. (\ref{21d}), we get
\begin{align}\label{24d}
m(t,r)=\frac{r^3\kappa T^0_0}{6f_R}-\frac{1}{6}\int^r_0\frac{r^3}{f_R}\left[\kappa(T^0_0)'+\frac{3\bar{\tau}_0}{ r}\right]dr.
\end{align}
It is clear from the above expression that the interior geometric mass of the solid sphere depends upon the energy density along with the curvature emendations resulting from the Palatini's gravitational technique. Due to this reason, our result is different from the mass function appearing in Eq. (47) in \cite{41}. The components of stress-energy tensor assumes the values as $T^0_0=\tilde{\rho}$ and
$T^2_2-T^1_1=\Pi$ for the following three particular situations:

\textbf{Static Regime}

(i) In this situation, all the time derivatives and the velocity $\omega$ vanish.

\textbf{Quasi-static Regime}

(ii) In this situtaion, we assume $\omega^2\approx\dot{\nu}\dot{\lambda}\approx\ddot{\nu} \approx\dot{\omega}\approx\ddot{\lambda} \approx\dot{\lambda}^2\approx0$.

\textbf{Right after the withdrawal from Equilibrium State}

(iii) In such a situation, we suppose that $\dot{\omega}\neq0$ but $\dot{\lambda}\approx\omega\approx\dot{\nu}\approx0$.\\
Bearing in mind these particular cases, we work out Eqs. (\ref{23d}) and (\ref{24d}) again as follows
\begin{align}\nonumber
E &=-\frac{1}{2r^3}\int^r_0 \frac{r^3}{f_R} \left(\kappa\tilde{\rho}'+\frac{3\bar{\tau}_0}{r}\right)dr+\frac{\kappa \Pi}{2f_R}+\frac{1}{2f_R}(\bar{\tau}_0+\bar{\tau}_1),\\\nonumber
m(t,r)&=\frac{ r^3 \kappa\tilde{\rho}}{6f_R}-\frac{1}{6}\int^r_0\frac{r^3}{f_R}\left(\kappa\tilde{\rho}'+\frac{3\bar{\tau}_0}{ r}\right)dr.
\end{align}
It is noteworthy that $E$ is linked with the physical fluid characteristics (anisotropy and density homogeneity in this case) whereas the mass function is connected with homogeneous energy density along with the change produced by inhomogeneity in the value for energy density.

\subsection{The Active Gravitational Spherical Mass}

The total energy inside the fluid content can be well-defined using an expression that incorporates fluid variables via stress-energy tensor components. We call it the active gravitational mass of the astronomical object under consideration. Mathematically, we write it as
\begin{align}\nonumber
m_T&=\frac{1}{2}\int^{r_\Sigma}_0
e^{(\nu+\lambda)/2}r^2\kappa(\left.T^0_0\right.^{eff}-\left.T^1_1\right.^{eff}-2\left.T^2_2\right.^{eff})dr+\frac{1}{2}\int^{r_\Sigma}_0
e^{(\nu+\lambda)/2}r^2 f_R(R_0)\\\nonumber &\times
\frac{\partial}{\partial t}\left[\frac{\partial
L}{\frac{\partial}{\partial t}[\partial(g^{\alpha\sigma}
\sqrt{-g})]}\right]g^{\alpha\sigma}dr.
\end{align}
Here, $L$ symbolize the Lagrangian density. Entirely inside the boundary $\Sigma$, the mass of solid sphere having radius $r$ can be defined as
\begin{align}\nonumber
m_T&=\frac{1}{2}\int^r_0
e^{(\nu+\lambda)/2}r^2\kappa(\left.T^0_0\right.^{eff}-\left.T^1_1\right.^{eff}-2\left.T^2_2\right.^{eff})dr+\frac{1}{2}\int^r_0
e^{(\nu+\lambda)/2}r^2 f_R(R_0)\\\nonumber &\times
\frac{\partial}{\partial t}\left[\frac{\partial
L}{\frac{\partial}{\partial t}[\partial(g^{\alpha\sigma}
\sqrt{-g})]}\right]g^{\alpha\sigma}dr.
\end{align}
Putting into use the Palatini gravitational equational together with the definition of interior spherical mass, we infer
\begin{equation}\label{25d}
m_T= e^{\nu+\lambda/2}f_R\left[m(t,r)-\frac{ r^3}{2f_R}\left(\kappa T^1_1-\bar{\tau}_1\right)\right]-\frac{1}{2}\int^r_0 e^{\lambda-\nu/2}\dot{\lambda} \dot{f_R} dr.
\end{equation}
Putting the value of component $T^1_1$ from Eq. (\ref{5d}) and $m$ from Eq. (\ref{18d}), we attain
\begin{equation}\label{26d}
m_T=\frac{r^2 \nu' e^{\frac{\nu-\lambda}{2}}f_R}{2}-\int r^2\dot{f_R}\dot{\lambda}e^{\frac{\nu-\lambda}{2}} dr.
\end{equation}
For a test particle initially at rest, the static field induces the following value for the acceleration
\begin{equation}\nonumber
a=\sqrt{\frac{m_T e^{-\nu/2}}{r^2}-\frac{e^{\frac{-\nu-\lambda}{2}}f_R'^2}{4f_R^2}}.
\end{equation}
Taking the radial derivative of Eq. (\ref{26d}) and using it together with Eqs. (\ref{21d}) and (\ref{25d}), we can write
\begin{align}\nonumber
&-3m_T +rm_T'=\left(\frac{\nu' f_R'}{f_R}-\frac{\dot{\lambda}\dot{f_R}}{f_R}\right)\times\frac{r^3 f_Re^{\frac{\nu-\lambda}{2}}}{2}
+\left(\frac{\ddot{\lambda}}{2}+\frac{\dot{\lambda}^2}{4}-\frac{\dot{\lambda}\dot{\nu}}{4}\right)\times r^3f_Re^{\frac{\lambda-\nu}{2}}+\frac{r^3\kappa e^{\frac{\nu+\lambda}{2}}}{2}\times \\\nonumber &\left(T^0_0-T^1_1-2T^2_2-\frac{\bar{\tau}_0}{\kappa}+\frac{\bar{\tau}_1}{\kappa}+\frac{2\bar{\tau}_2}{\kappa}\right)
-r^3e^{\frac{\nu+\lambda}{2}}\left[\kappa \left(\frac{T^0_0}{2f_R}-\frac{T^0_0}{2f_R}-\frac{T^0_0}{2f_R}+ \frac{\bar{\tau}_1+\bar{\tau}_2-\bar{\tau}_0}{\kappa}\right)\right.\\\nonumber & \left. +\frac{E}{f_R}\right]+\frac{3}{2}\int^r_0 r^2 e^{\frac{\lambda-\nu}{2}}\dot{\lambda} \dot{f_R}dr.
\end{align}
Employing integration w.r.t $r$, we get
\begin{align}\nonumber
m_T&= (m_T)_\Sigma\left(\frac{r}{r_\Sigma}\right)^3-r^3 \int_r^{r_\Sigma}\frac{e^{\frac{\nu-\lambda}{2}}}{2r}\left(\nu' f_R'-\dot{\lambda}\dot{f_R}\right)dr-r^3\int_r^{r_\Sigma}r^2f_Re^{\frac{\lambda-\nu}{2}}\left(\frac{\ddot{\lambda}}{2}
+\frac{\dot{\lambda}^2}{4}-\frac{\dot{\lambda}\dot{\nu}}{4}\right)\\\label{27d}&dr
-r^3\int^{r_\Sigma}_r\frac{ e^{\frac{\nu+\lambda}{2}}}{r} \left(\frac{\kappa}{2}(T^1_1-T^2_2-\frac{\bar{\tau}_2-\bar{\tau}_1}{\kappa})
-\frac{E}{f_R}\right)dr-r^3 \int_r^{r_\Sigma}\frac{3}{2r^4}\int^r_0 r^2 e^{\frac{\lambda-\nu}{2}}\dot{\lambda} \dot{f_R}drdr.
\end{align}
Previously in literature, as given in Eq. (57) in \cite{41}, the active gravitational mass in the frame of Einstein's GR was found to be dependent only on the fluid variables along with the derivatives of metric coefficients. But in our theory ($f(R)$ by Palatini's approach), it comes out to be dependent also on the dark source terms which means that the dark source also affects the strength of an astronomical body's gravitational flux.
Eq. (\ref{23d}) when utilized here, provides the following outcome
\begin{align}\nonumber
m_T&= (m_T)_\Sigma\left(\frac{r}{r_\Sigma}\right)^3-r^3 \int_r^{r_\Sigma}\frac{e^{\frac{\nu-\lambda}{2}}}{2r}\left(\nu' f_R'-\dot{\lambda}\dot{f_R}\right)dr-r ^3\int_r^{r_\Sigma}\frac{f_Re^{\frac{\lambda-\nu}{2}}}{r}\left(\frac{\ddot{\lambda}}{2}+\frac{\dot{\lambda}^2}{4}-\frac{\dot{\lambda}\dot{\nu}}{4}\right)
dr\\\nonumber&-r^3\int^{r_\Sigma}_r\frac{ e^{\frac{\nu+\lambda}{2}}}{r} \left(\kappa T^1_1-\kappa T^2_2+\bar{\tau}_2-\bar{\tau}_1+\frac{\kappa f_R}{2r^3}\int_0^r \frac{r^3}{f_R}\left((T^0_0)'+\frac{3\bar{\tau}_0}{\kappa r}\right)dr
\right)dr.
\end{align}

\subsection{Analysis of Structure Formation and Evolutionary Phases of Spherical Star}

This section focuses on the manipulation of ceratin differential equations, each one explicating the physical phenomena that the spherical object confronts. In this regard, the first conservation equation is inferred as
\begin{align}\nonumber
&\tilde{\rho}^*+(\tilde{\rho}+\tilde{P}_r)\left(\theta-\frac{2\dot{f_R}e^{-\nu/2}}{f_R (1-\omega^2)^{1/2}}-\frac{2f_R'\omega e^{-\lambda/2}}{f_R (1-\omega^2)^{1/2}}\right)-\frac{2}{3}\left(\theta+\frac{\sigma}{2}-\frac{2\dot{f_R}e^{-\nu/2}}{f_R (1-\omega^2)^{1/2}}\right.\\\nonumber &\left.-\frac{3f_R'\omega
e^{-\lambda/2}}{f_R (1-\omega^2)^{1/2}}+\frac{3\dot{f_R}e^{-\nu/2}(1-\omega^2)^{1/2}}{2\omega^2 f_R}\right)\Pi+
\tilde{q}^\dagger+2\tilde{q}\left[\sqrt{a^2-\frac{f_R'^2\omega^2 e^{-\lambda}}{4f_R^2(1-\omega^2)^2}+\varepsilon_1}\right.\\\nonumber &\left.+\frac{s^1}{r}\right]+\varphi_1^{(D)}=0,
\end{align}
with the value
\begin{align}\nonumber
\varphi_1^{(D)}&= \mathcal{T}_{0,1}+\mathcal{T}^0_{0,0}+\mathcal{T}^0_0 \left[\frac{\dot{\lambda}}{2}+\frac{\dot{f_R}}{2f_R}\right]+\mathcal{T}^1_0 \left[\frac{\lambda'}{2}+\frac{2}{r}+\frac{3f_R'}{2f_R}\right]-\mathcal{T}^0_1\left[\frac{\nu' e^{\nu-\lambda/2}}{2}+\frac{f_R' e^{\nu-\lambda/2}}{2f_R}\right]\\\nonumber&+T^0_0\left[\frac{\dot{f_R}}{2f_R}\right]+T^1_0
\left[\frac{3f_R'}{2f_R}\right]-\mathcal{T}^1_1\left[\frac{\dot{\lambda}}{2}+\frac{\dot{f_R}}{2f_R}\right]-T^1_1\left[\frac{\dot{f_R}}{2f_R}\right]-T^0_1\left[\frac{f_R' e^{\nu-\lambda/2}}{2f_R}\right].
\end{align}
The second conservation equation is inferred as
\begin{align}\nonumber
&\tilde{P}_r^\dagger+(\tilde{\rho}+\tilde{P}_r)\sqrt{a^2-\frac{f_R'^2 \omega^2 e^{-\lambda}}{4f_R^2 (1-\omega^2)}+\varepsilon_1}+\frac{2s^1}{r}\Pi= \frac{\tilde{q}}{3}\left[\sigma-4\theta+\frac{6\omega f_R' e^{-\lambda/2}}{f_R \sqrt{1-\omega^2}}\right.\\\nonumber &\left.+\frac{8\dot{f_R}e^{-\nu/2}}{f_R \sqrt{1-\omega^2}}+\frac{3\dot{f_R}e^{-\nu/2}\sqrt{1-\omega^2}}{f_R \omega^2}\right]+\tilde{q}^\star-\varphi_2^{(D)}.
\end{align}
where
\begin{align}\nonumber
\varphi_2^{(D)}&=\mathcal{T}^1_{1,1}+\mathcal{T}^0_{0,0}+\mathcal{T}^1_1 \left(\frac{\nu'}{2}+\frac{f_R'}{2f_R}\right)-\mathcal{T}^0_0\left(\frac{\nu'}{2}+\frac{f_R'}{2f_R}\right)+\tilde{T}^1_1\left(\frac{2}{r}+\frac{f_R'}{f_R}\right)
-\mathcal{T}^1_0 \left(\frac{\dot{\lambda}e^{\lambda-\nu}}{2}\right.\\\nonumber &\left.+\frac{\dot{f_R}e^{\lambda-\nu}}{2f_R}\right) -\frac{2\mathcal{T}^2_2}{r}+\mathcal{T}^0_1\left(\frac{\dot{\nu}}{2}+\frac{\dot{f_R}}{2f_R}\right).
\end{align}
Both of the above equations when compared with Eqs. (59) and (60) in \cite{41} exhibit that the presence of dark source produces significant effects on the stress-energy conservation of astronomical systems.
The Raychaudhuri equation that explicates the flow kinematics induces the following outcome
\begin{align}\nonumber
&\left[\theta-\frac{2\dot{f_R}e^{-\nu/2}}{f_R (1-\omega^2)^{1/2}}-\frac{2f_R'\omega e^{-\lambda/2}}{f_R (1-\omega^2)^{1/2}}
\right]^\star +\frac{1}{3}\left[\theta-\frac{2\dot{f_R}e^{-\nu/2}}{f_R (1-\omega^2)^{1/2}}-\frac{2f_R'\omega e^{-\lambda/2}}{f_R (1-\omega^2)^{1/2}}
\right]^2\\\nonumber &+\frac{1}{6}\left[\sigma+\frac{3\dot{f_R}e^{-\nu/2}(1-\omega^2)^{1/2}}{f_R \omega^2}-\frac{2f_R'\omega
e^{-\lambda/2}}{f_R (1-\omega^2)^{1/2}}\right]^2-\left[\sqrt{a^2-\frac{f_R'^2 \omega^2 e^{-\lambda}}{4f_R^2 (1-\omega^2)}+\varepsilon_1}\right]^\dagger-\\\nonumber &\left[a^2-\frac{f_R'^2 \omega^2 e^{-\lambda}}{4f_R^2 (1-\omega^2)}+\varepsilon_1\right]
-\frac{2s^1}{r}\sqrt{a^2-\frac{f_R'^2 \omega^2 e^{-\lambda}}{4f_R^2 (1-\omega^2)}+\varepsilon_1}=-\frac{\kappa}{2}(\tilde{\rho}+3\tilde{P}_r)+\kappa\Pi,
\end{align}
which being dissimilar to Eq. (61) in \cite{41} for GR, demonstrates that the dark source also plays its part in characterizing the fluid flow in the spacetime background. The Ricci identities yield
\begin{align}\nonumber
&\left[\frac{\sigma}{2}+\theta+\frac{3\dot{f_R}e^{-\nu/2}(1-\omega^2)^{1/2}}{2\omega^2f_R}-\frac{3\omega f_R' e^{-\lambda/2}}{f_R (1-\omega^2)^{1/2}}-\frac{2\dot{f_R}e^{-\nu/2}}{f_R (1-\omega^2)^{1/2}}
\right]^\dagger=-\frac{3s^1}{2r} \\\nonumber &\times\left(\sigma+\frac{3\dot{f_R}e^{-\nu/2}(1-\omega^2)^{1/2}}{\omega^2 f_R}-\frac{2\omega f_R' e^{-\lambda/2}}{f_R (1-\omega^2)^{1/2}}\right)+\frac{3}{2}\kappa \tilde{q},
\end{align}
which, unlike Eq. (62) in \cite {41}, signify that the fluid dynamics described by the shear and expansion scalar and the heat dissipation factor is also influenced by the inclusion of dark source in the gravitational theory. From Eqs. (\ref{3d}), (\ref{16d}) together with Ricci identities produce
\begin{align}\nonumber
&-E-\frac{\kappa\Pi}{2}=\left[\left(a^2-\frac{f_R'^2 \omega^2 e^{-\lambda}}{4f_R^2 (1-\omega^2)}+\varepsilon_1\right)^{1/2}\right]^\dagger +\left[a^2-\frac{f_R'^2 \omega^2 e^{-\lambda}}{4f_R^2 (1-\omega^2)}+\varepsilon_1\right]\\\nonumber &+\frac{1}{2}\left[\sigma+\frac{3\dot{f_R}e^{-\nu/2}(1-\omega^2)^{1/2}}{\omega^2f_R}-\frac{2\omega f_R' e^{-\lambda/2}}{f_R (1-\omega^2)^{1/2}}\right]^\star+\frac{1}{3}\left[\left(\theta-\frac{2\dot{f_R}e^{-\nu/2}}{f_R (1-\omega^2)^{1/2}}-\frac{2f_R'\omega e^{-\lambda/2}}{f_R (1-\omega^2)^{1/2}}\right)\right. \\\nonumber&\times\left.\left(\sigma+\frac{3\dot{f_R}e^{-\nu/2}(1-\omega^2)^{1/2}}{\omega^2f_R}-\frac{2\omega f_R' e^{-\lambda/2}}{f_R (1-\omega^2)^{1/2}}\right)\right]-\frac{s^1}{r}\sqrt{a^2-\frac{f_R'^2 \omega^2 e^{-\lambda}}{4f_R^2 (1-\omega^2)}+\varepsilon_1}\\\nonumber &+\frac{1}{12}\left[\sigma+\frac{3\dot{f_R}e^{-\nu/2}(1-\omega^2)^{1/2}}{\omega^2f_R}-\frac{2\omega f_R' e^{-\lambda/2}}{f_R (1-\omega^2)^{1/2}}\right].
\end{align}
From the comparison of this outcome with Eq. (63) in \cite{41}, it is worth-observing that the incorporation of dark source quantities (emerging because of Palatini's formalism) influences the tidal force and the anisotropy factor for the astronomical system.
The Bianchi identities in combination with Weyl tensor produce following two differential equations
\begin{align}\nonumber
&\left(\frac{\kappa \tilde{P}_r}{2f_R}+\frac{3m}{r^3}\right)\left[\theta-\frac{2\dot{f_R}e^{-\nu/2}}{f_R (1-\omega^2)^{1/2}}-\frac{3f_R'\omega e^{-\lambda/2}}{f_R (1-\omega^2)^{1/2}}+\frac{\sigma}{2}+\frac{3\dot{f_R}e^{-\nu/2}(1-\omega^2)^{1/2}}{\omega^2f_R}\right]\\\nonumber &+\left[E-\frac{\kappa\Pi}{2}+\frac{\kappa \tilde{\rho}}{2}\right]^\star=-\frac{3s^1}{2r}\kappa\tilde{q},
\end{align}
and
\begin{align}\nonumber
&\left[E-\frac{\kappa\Pi}{2}+\frac{\kappa \tilde{\rho}}{2}\right]^\dagger=\frac{3s^1}{r}\left(\frac{\kappa\Pi}{2}-E
 \right)+\frac{\kappa \tilde{q}}{2}\left[\frac{\sigma}{2}+\theta+\frac{3\dot{f_R}e^{-\nu/2}(1-\omega^2)^{1/2}}{2\omega^2f_R}\right.\\\nonumber &\left.-\frac{3\omega f_R' e^{-\lambda/2}}{f_R (1-\omega^2)^{1/2}}-\frac{2\dot{f_R}e^{-\nu/2}}{f_R (1-\omega^2)^{1/2}}
\right],
\end{align}
each of which shows a physical meaning distinct from that provided by Eqs. (64,65) in \cite{41}. This particular distinction is because we are taking the emendations in Ricci scalar curvature which, in turn, has peculiar effects on the fluid flow.
Equation (\ref{19d}) can be reformed as
\begin{equation}\nonumber
\frac{3m}{r^3}=\frac{E}{f_R}-\frac{\kappa\Pi}{2f_R}+\frac{\kappa \tilde{\rho}}{2f_R}-\frac{\bar{\tau}_0+\bar{\tau}_1}{2f_R}.
\end{equation}

\section{Breaking-up the Riemann Tensor via Orthogonal Decomposition Method}

Following the approach used by Bel \cite{40}, we introduce three explicit tensors $X_{\varrho\beta}, Y_{\varrho\beta}$ and $Z_{\varrho\beta}$ as
\begin{eqnarray}\nonumber
Y_{\varrho\beta}=R_{\varrho\gamma\beta\delta}u^\gamma
u^\delta,\quad
Z_{\varrho\beta}=^*R_{\varrho\gamma\beta\delta}u^\gamma
u^\delta=\frac{1}{2}\eta_{\varrho\gamma\epsilon\rho}R^{\epsilon\rho}_{\beta\delta}u^\gamma
u^\delta,\quad
X_{\varrho\beta}=^*R^*_{\varrho\gamma\beta\delta}u^\gamma
u^\delta=\frac{1}{2}\eta_{\varrho\gamma}^{\epsilon\rho}R^*_{\epsilon\rho\beta\delta}u^\gamma
u^\delta,
\end{eqnarray}
with the value
\begin{equation}\nonumber
R^*_{\varrho\beta\gamma\delta}=\frac{1}{2}\eta_{\epsilon\omega\gamma\delta}R^{\epsilon\omega}_{\varrho\beta}.
\end{equation}
Putting into use the Palatini $f(R)$ field equations given in Eq. (\ref{3d}), we infer
\begin{align}\nonumber
R^{\sigma\alpha}_{\pi\beta}=C^{\sigma\alpha}_{\pi\beta}+2\kappa
\left.T^{(eff)}\right.^{[\sigma}_{[\pi}\delta^{\alpha]}_{\beta]}+\kappa
\left.T^{(eff)}\right.\left(\frac{1}{3}\delta^\sigma_{[\pi}\delta^\alpha_{\beta]}-\delta^{[\sigma}_{[\pi}\delta^{\alpha]}_{\beta]}
\right).
\end{align}
The Riemann tensor is broken up into three distinct parts given as under
\begin{equation}\nonumber
R^{\sigma\alpha}_{\pi\beta}=R^{\sigma\alpha}_{(I)\pi\beta}+R^{\sigma\alpha}_{(II)\pi\beta}+R^{\sigma\alpha}_{(III)\pi\beta},
\end{equation}
where each part holds the following value
\begin{align}\nonumber
R^{\sigma\alpha}_{(I)\pi\beta}=&2\kappa\tilde{\rho}u^{[\sigma}u_{[\pi}\delta^{\alpha]}_{\beta]}-2\kappa \hat{P}
h^{[\sigma}_{[\pi}\delta^{\alpha]}_{\beta]}+\left(\kappa
\tilde{\rho}-3\kappa\check{P}\right.\\\nonumber&\left.+2f-2Rf_R-3\Box
f_R+\frac{3(\nabla f_R)^2}{f_R}-\frac{3\nabla^\gamma f_R\nabla_\gamma f_R}{f_R}\right)\left(\frac{1}{3}\delta^\sigma_{[\pi}
\delta^\alpha_{\beta]}-\delta^{[\sigma}_{[\pi}\delta^{\alpha]}_{\beta]}\right)
\\\nonumber &+2 \left[\nabla^{[\sigma}\nabla_{[\pi}f_R \delta^{\alpha]}_{\beta]}-\Box f_R \delta^{[\sigma}_{[\pi}\delta^{\alpha]}_{\beta]}+(f-Rf_R)\delta^{[\sigma}_{[\pi}\delta^{\alpha]}_{\beta]}
+\frac{3}{4f_R}(\delta^{[\sigma}_{[\pi}\delta^{\alpha]}_{\beta]})(\nabla f_R)^2\right.\\\nonumber& \left.
-\frac{3}{2f_R}(\nabla^{[\sigma}f_R \nabla _{[\pi} f_R \delta^{\alpha]}_{\beta]})\right],\\\nonumber R^{\sigma\alpha}_{(II)\pi\beta}=&2\kappa \Pi\left[s^{[\sigma} s_{[\pi}\delta^{\alpha]}_{\beta]} +\frac{1}{3} h^{[\sigma}_{[\pi}\delta^{\alpha]}_{\beta]}\right]+2\kappa\left(\tilde{q}s^{[\sigma}
u_{[\pi}\delta^{\alpha]}_{\beta]}+\tilde{q}u^{[\sigma}s_{[\pi}\delta^{\alpha]}_{\beta]}\right),\\\nonumber
R^{\sigma\alpha}_{(III)\pi\beta}=&4u^{[\sigma}u_{[\pi}E^{\alpha]}_{\beta]}-\epsilon^{\sigma\alpha}_\mu
\epsilon_{\pi\beta\nu}E^{\mu\nu}.
\end{align}
Such a breakdown of the Riemann tensor was first proposed by Herrera in \cite{41} for GR. Eqs. (73) of this article coincide with our results if we consider the restrictions that $f_{R}\rightarrow 1, f(R)=R$ and all the derivatives of $f_R$ equal to zero.
Expressions for tensorial quantities described above take the following values
\begin{align}\label{28d}
X_{\varrho\beta}&=\frac{\kappa h_{\varrho\beta}}{3}\tilde{\rho}
-E_{\varrho\beta}+\frac{\kappa}{2} \Pi_{\varrho\beta}+\chi_{\varrho\beta},\\\label{29d}
Y_{\varrho\beta}&=\frac{\kappa h_{\varrho\beta}}{6}\tilde{\rho}+\frac{\kappa h_{\varrho\beta}}{2}\hat{P}+\frac{\kappa}{2} \Pi_{\varrho\beta}+E_{\varrho\beta}+\psi_{\varrho\beta},\\\label{30d} Z_{\varrho\beta}&= \frac{\kappa \tilde{q}}{2}(s^\rho \epsilon_{\varrho\rho\beta})
+\tilde{\xi}_{\varrho\beta},
\end{align}
with the values of curvature emendation terms as follows
\begin{align}\nonumber
\chi_{\varrho\beta}&=\frac{g_{a\varrho}\epsilon^{\epsilon\rho a}}{8}\left[(\nabla ^\pi \nabla_\epsilon f_R-\frac{3}{2f_R}\nabla ^\pi f_R \nabla_\epsilon f_R)\epsilon_{\beta\pi\rho}+(\frac{3}{2f_R}\nabla ^\pi f_R \nabla_\rho f_R-\nabla ^\pi \nabla_\rho f_R)\epsilon_{\beta\pi\epsilon}\right.\\\nonumber &\left.+(\frac{3}{2f_R}\nabla ^\theta f_R \nabla_\epsilon f_R-\nabla ^\theta \nabla_\epsilon f_R)\epsilon_{\theta\beta\rho}+(\nabla ^\theta \nabla_\rho f_R-\frac{3}{2f_R}\nabla ^\theta f_R \nabla_\rho f_R)\epsilon_{\theta\beta\epsilon} \right]+\frac{f-Rf_R}{6}h_{\varrho\beta}\\\nonumber &+
\frac{(\nabla f_R)^2}{2f_R}h_{\varrho\beta}-\frac{\nabla^{\delta}f_R \nabla_{\delta}f_R}{2f_R}h_{\varrho\beta},\\\nonumber
\psi_{\varrho\beta}&=\frac{1}{2}\left(\nabla_\varrho \nabla_\beta f_R-u_\beta u^\delta\nabla_\varrho\nabla_\delta f_R-u_\varrho u_\theta\nabla^\theta\nabla_\beta f_R-u_\theta u^\delta g_{\varrho\beta}\nabla^\theta\nabla_\delta f_R\right)+\frac{3}{2f_R}[-\nabla_\varrho f_R \nabla_\beta f_R+\\\nonumber &(\nabla_\varrho f_R \nabla_\delta f_R)u_\beta u^\delta+(\nabla^\theta f_R \nabla_\beta f_R)u_\varrho u_\theta-(\nabla^\theta f_R \nabla_\delta f_R)u_\theta u^\delta g_{\varrho\beta}]
-\left(\frac{f-Rf_R}{6}\right)h_{\varrho\beta}-\\\nonumber &\frac{(\nabla f_R)^2}{4f_R}h_{\varrho\beta}+\frac{\nabla^\beta f_R \nabla_\beta f_R}{2}h_{\varrho\beta},\\\nonumber \tilde{\xi}_{\varrho\beta}&=-\frac{1}{2}[(\nabla^\gamma \nabla_\delta f_R)u^\delta \epsilon_{\varrho\gamma\beta}]+\frac{3}{4f_R}(\nabla^\gamma \nabla_\delta f_R)u^\delta \epsilon_{\varrho\gamma\beta}.
\end{align}

\subsection{Super-Poynting Vector and Bel Superenergy}

The values for Bel superenergy and Super-Poynting vector turn out to be as under
\begin{align}\nonumber
&\bar{W}=\frac{5\kappa^2 \tilde{\rho}^2}{24}+\frac{\kappa^2 \Pi^2}{6}+\frac{2E^2}{3}+\frac{\kappa^2 \tilde{\rho}\hat{P}}{4}
+\frac{3\kappa^2 \hat{P}}{8}+\frac{\kappa^2 \tilde{q}^2}{2}+\frac{M_1+M_2}{2}+M_3,\\\nonumber
&\bar{P}_\beta=\frac{\kappa^2 \tilde{q}}{2}\left(\tilde{\rho}+P_r\right)s_\beta-\varsigma_\beta,
\end{align}
where
\begin{align}\nonumber
\varsigma_\beta&=\frac{\kappa \tilde{q}}{2}(s_\alpha h^\delta_\beta-s_\beta\delta^\delta_\alpha+u_\alpha u^\delta s_\beta)\varphi^\alpha_\delta
+\xi^{\gamma\delta}\epsilon_{\beta\gamma\alpha}\left[E^\alpha_\delta+\frac{\kappa h^\alpha_\delta}{6}\tilde{\rho}+\frac{\kappa h^\alpha_\delta}{2}\hat{P}+\frac{\kappa}{2} \Pi^\alpha_\delta+\varphi^\alpha_\delta\right]\\\nonumber&+\xi^{\gamma\delta}\epsilon_{\beta\gamma\alpha}\varphi^\alpha_\delta+ \xi^{\gamma\delta}\epsilon_{\beta\gamma\alpha}\left[-E^\alpha_\delta+\frac{\kappa h^\alpha_\delta}{3}\tilde{\rho}
+\frac{\kappa}{2}\Pi^\alpha_\delta+\vartheta^\alpha_\delta\right]+\frac{\kappa \tilde{q}}{2}(s_\alpha h^\delta_\beta-s_\beta\delta^\delta_\alpha+u_\alpha u^\delta s_\beta)\vartheta^\alpha_\delta\\\nonumber &+\xi^{\gamma\delta}\epsilon_{\beta\gamma\alpha}\vartheta^\alpha_\delta.
\end{align}
The curvature emendations are as follows
\begin{align}\nonumber
M_1&= -E_{\mu\nu}\vartheta^{\mu\nu}+\frac{\kappa \tilde{\rho}h_{\mu\nu}\vartheta^{\mu\nu}}{3}+\frac{\kappa \Pi_{\mu\nu}\vartheta^{\mu\nu}}{2}-E^{\mu\nu}\vartheta_{\mu\nu}+\frac{\kappa \tilde{\rho}h^{\mu\nu}\vartheta_{\mu\nu}}{3}+\frac{\kappa \Pi^{\mu\nu}\vartheta_{\mu\nu}}{2}+\vartheta_{\mu\nu}\vartheta^{\mu\nu},\\\nonumber
M_2&=\frac{\kappa \tilde{\rho}h_{\mu\nu}\tilde{\vartheta}^{\mu\nu}}{6}+\frac{\kappa \hat{P} h_{\mu\nu}\tilde{\vartheta}^{\mu\nu}}{2}+E_{\mu\nu}\tilde{\vartheta}^{\mu\nu}+\frac{\kappa \Pi_{\mu\nu}\tilde{\vartheta}^{\mu\nu}}{2},\\\nonumber
M_3&=\frac{\kappa \tilde{q}s^{\rho} \epsilon_{\mu\rho\nu}\xi^{\mu\nu}}{2}+\frac{\kappa \tilde{q}}{2}s^{\rho}\epsilon^{\mu\nu}_{\rho}\xi_{\mu\nu}+\xi_{\mu\nu}\xi^{\mu\nu}.
\end{align}
The Bel Robinson tensor depicted as $W=E^{\mu\eta}E_{\mu\eta}=\frac{2E^2}{3}$ when subtracted form the Bel superenergy, the following outcome is established
\begin{equation}\nonumber
\bar{W}-W=\frac{5\kappa^2 \tilde{\rho}^2}{24}+\frac{\kappa^2 \tilde{\rho}\hat{P}}{4}
+\frac{3\kappa^2 \hat{P}}{8}+\frac{\kappa^2 \Pi^2}{6}+\frac{\kappa^2 \tilde{q}^2}{2}+\frac{M_1+M_2}{2}+M_3.
\end{equation}
\subsection{Evaluation of Modified Scalar Functions}
In this Section, five structure scalars are ascertained from the trace and trace-free parts of tensorial quantities $X_{\varrho\beta}, Y_{\varrho\beta}$ and $Z_{\varrho\beta}$. This analysis is carried out by following the work of Herrera \cite{41}. \\
Another technique of defining the tensor $X_{\varrho\beta}$ is
\begin{eqnarray}\nonumber
X_{\varrho\beta}=\frac{h_{\varrho\beta}}{3} Tr X+X_{<\varrho\beta>},\quad \textmd{where}\quad X_{<\varrho\beta>}=h^\gamma_\varrho h^\nu_\beta\left(X_{\varrho\beta}-\frac{h_{\gamma\nu}}{3} Tr X\right).
\end{eqnarray}
This leads us to the following outcome
\begin{equation}\nonumber
X_{<\varrho\beta>}=X_{TF}\left(s_\varrho s_\beta+\frac{h_{\varrho\beta}}{3}\right).
\end{equation}
Now computing the trace component i.e. $Tr X=X^\varrho_\varrho=X_T$ and the trace-free component $X_{TF}$, we infer
\begin{align}\label{31d}
X_T&=\kappa\tilde{\rho}+\eta_1,\\\label{32d}
X_{TF}&=\frac{\kappa \Pi}{2}-E+ \vartheta_1^{(D)},
\end{align}
where
\begin{align}\nonumber
\eta_1&=\frac{1}{4}\left[\left(\frac{3}{2f_R}\nabla^\nu f_R \nabla_\epsilon f_R-\nabla^\nu \nabla_\epsilon f_R\right)h^\epsilon_\nu+\left(\frac{3}{2f_R}
\nabla^\nu f_R \nabla_\rho f_R -\nabla^\nu \nabla_\rho f_R\right)h^\rho_\nu\right.\\\nonumber&\left.+\left(\frac{3}{2f_R}\nabla^\alpha f_R \nabla_\epsilon f_R-\nabla^\alpha \nabla_\epsilon f_R\right)h^\epsilon_\alpha+\left(\frac{3}{2f_R}\nabla^\alpha f_R \nabla_\rho f_R-\nabla^\alpha \nabla_\rho f_R\right)h^\rho_\alpha\right]+\left(\frac{f-Rf_R}{2}\right)\\\nonumber &+\frac{3(\nabla f_R)^2}{2f_R}-\frac{3\nabla^\gamma f_R \nabla_\gamma f_R}{2f_R}.
\end{align}
Eq. (\ref{31d} and (\ref{32d}) correspond to Eqs. (89, 91) in \cite{41} with a significant distinction that the influence of the dark source is incorporated in the trace and trace free parts of $X_{\varrho\beta}$.\\
In the same manner, the trace and trace-free components of $Y_{\varrho\beta}$ are acquired as
\begin{align}\label{33d}
Y_T&=\frac{\kappa}{2}\left[\tilde{\rho}+3\tilde{P}_r-2\Pi\right]+\eta_2,\\\label{34d} Y_{TF}&=\frac{\kappa\Pi}{2}+E+ \vartheta_2^{(D)},
\end{align}
where
\begin{align}\nonumber
\eta_2&= \frac{\Box f_R}{2}-\frac{3\Box f_R}{2f_R}+\left(\frac{3}{2f_R}\nabla^\nu f_R \nabla_\delta f_R-\frac{\nabla^\nu \nabla_\delta f_R}{2}\right)u_\nu u^\delta+\left(\frac{3}{2f_R}\nabla^\pi \nabla_\nu f_R-\frac{\nabla^\pi \nabla_\nu f_R}{2}\right)\\\nonumber &
u^\nu u_\pi +2\left(\nabla^\pi \nabla_\delta f_R -\frac{3\nabla^\pi f_R \nabla_\delta f_R}{f_R}\right)u_\pi u^\delta- \left(\frac{f- Rf_R}{2}\right)-\frac{3(\nabla f_R)^2}{4f_R}+\frac{3\nabla^\gamma f_R \nabla_\gamma f_R}{2f_R}.
\end{align}
Again, on comparison of Eq. (\ref{33d}) and (\ref{34d} )with Eqs. (92, 94) of \cite{41}, it is worthy to note that the trace and trace-free components of $Y_{\varrho\beta}$ are also influenced by the existence of dark source.\\
From the tensorial quantity $Z_{\varrho\beta}$ connected with the heat dissipative flux, we infer the last scalar $Z$ as
\begin{equation}\label{35d}
Z^2=Z_{\varrho\beta}Z^{\varrho\beta}=\frac{\kappa^2 \tilde{q}^2}{2}+\tilde{\xi},
\end{equation}
with $\vartheta_1^{(D)}=\frac{\chi_{\varrho\beta}}{s_\varrho s_\beta+\frac{h_{\varrho\beta}}{3}}$, $\vartheta_2^{(D)}=\frac{\psi_{\varrho\beta}}{s_\varrho s_\beta+\frac{h_{\varrho\beta}}{3}}$ and $\tilde{\xi}$ being equal to
\begin{equation}\nonumber
\tilde{\xi}=\frac{\kappa \tilde{q}}{2}s^\nu\epsilon_{\mu\nu\alpha}\xi^{\mu\alpha}+\frac{\kappa \tilde{q}}{2}s^\nu\epsilon^{\mu\alpha}_\nu\xi_{\mu\alpha}+\xi^{\mu\alpha}\xi_{\mu\alpha}.
\end{equation}

\subsection{Structure Formation and Evolution Equations in terms of Structure Scalars}

The stellar equations stated in Section \textbf{2.4} are restated in terms of five modified structure scalars as under
\begin{align}\nonumber
&\frac{\kappa\tilde{\rho}^*}{2}+\frac{1}{3}\left[X_T+Y_T+X_{TF}+Y_{TF}-\eta_1-\eta_2-\vartheta_1^{(D)}-\vartheta_2^{(D)}\right]
\left(\theta-\frac{2\dot{f_R}e^{-\nu/2}}{f_R
(1-\omega^2)^{1/2}}\right.\\\nonumber &\left.-\frac{2f_R'\omega
e^{-\lambda/2}}{f_R
(1-\omega^2)^{1/2}}\right)-\frac{1}{3}\left(\theta+\frac{\sigma}{2}-\frac{2\dot{f_R}e^{-\nu/2}}{f_R
(1-\omega^2)^{1/2}}-\frac{3f_R'\omega e^{-\lambda/2}}{f_R
(1-\omega^2)^{1/2}}+\frac{3\dot{f_R}e^{-\nu/2}(1-\omega^2)^{1/2}}{2\omega^2
f_R}\right)\\\nonumber &\times
\left(X_{TF}+Y_{TF}-\vartheta_1^{(D)}-\vartheta_2^{(D)}\right)+
\frac{\sqrt{2}}{2}\left[\sqrt{Z^2-\tilde{\xi}}\right]^\dagger+2\sqrt{Z^2-\tilde{\xi}}\\\label{36d}&\times\left[\sqrt{a^2-\frac{f_R'^2\omega^2
e^{-\lambda}}{4f_R^2(1-\omega^2)^2}+\varepsilon_1}+\frac{s^1}{r}\right]+\varphi_1^{(D)}=0,\\\nonumber
&\frac{\kappa\tilde{P}_r}{2}^\dagger+\frac{1}{3}\left[X_T+Y_T+X_{TF}+Y_{TF}-\eta_1-\eta_2-\vartheta_1^{(D)}-\vartheta_2^{(D)}\right]\sqrt{a^2-\frac{f_R'^2
\omega^2 e^{-\lambda}}{4f_R^2
(1-\omega^2)}+\varepsilon_1}\\\nonumber
&+\frac{s^1}{r}\left(X_{TF}+Y_{TF}-\vartheta_1^{(D)}-\vartheta_2^{(D)}\right)-
\frac{1}{3}\sqrt{2
Z^2-2\tilde{\xi}}\left[\frac{\sigma}{2}-2\theta+\frac{3\omega f_R'
e^{-\lambda/2}}{f_R \sqrt{1-\omega^2}}\right.\\\label{37d}
&\left.+\frac{4\dot{f_R}e^{-\nu/2}}{f_R
\sqrt{1-\omega^2}}+\frac{3\dot{f_R}e^{-\nu/2}\sqrt{1-\omega^2}}{f_R
\omega^2}\right]+\frac{\sqrt{2}}{2}\left[\sqrt{Z^2-\tilde{\xi}}\right]^\star+\varphi_2^{(D)}=0,\\\nonumber
&\left[\theta-\frac{2\dot{f_R}e^{-\nu/2}}{f_R
(1-\omega^2)^{1/2}}-\frac{2f_R'\omega e^{-\lambda/2}}{f_R
(1-\omega^2)^{1/2}} \right]^\star
+\frac{1}{3}\left[\theta-\frac{2\dot{f_R}e^{-\nu/2}}{f_R
(1-\omega^2)^{1/2}}-\frac{2f_R'\omega e^{-\lambda/2}}{f_R
(1-\omega^2)^{1/2}} \right]^2\\\nonumber
&+\frac{1}{6}\left[\sigma+\frac{3\dot{f_R}e^{-\nu/2}(1-\omega^2)^{1/2}}{f_R
\omega^2}-\frac{2f_R'\omega e^{-\lambda/2}}{f_R
(1-\omega^2)^{1/2}}\right]^2-\left[\sqrt{a^2-\frac{f_R'^2 \omega^2
e^{-\lambda}}{4f_R^2
(1-\omega^2)}+\varepsilon_1}\right]^\dagger-\\\label{38d}
&\left[a^2-\frac{f_R'^2 \omega^2 e^{-\lambda}}{4f_R^2
(1-\omega^2)}+\varepsilon_1\right]-\frac{2s^1}{r}\left[\left(a^2-\frac{f_R'^2
\omega^2 e^{-\lambda}}{4f_R^2
(1-\omega^2)}+\varepsilon_1\right)^{1/2}\right]=-Y_T
+\eta_2,\\\nonumber
&\left[\frac{\sigma}{2}+\theta+\frac{3\dot{f_R}e^{-\nu/2}(1-\omega^2)^{1/2}}{2\omega^2f_R}-\frac{3\omega
f_R' e^{-\lambda/2}}{f_R
(1-\omega^2)^{1/2}}-\frac{2\dot{f_R}e^{-\nu/2}}{f_R
(1-\omega^2)^{1/2}} \right]^\dagger=-\frac{3s^1}{2r} \\\label{39d}
&\times\left(\sigma+\frac{3\dot{f_R}e^{-\nu/2}(1-\omega^2)^{1/2}}{\omega^2
f_R}-\frac{2\omega f_R' e^{-\lambda/2}}{f_R
(1-\omega^2)^{1/2}}\right)+\frac{3}{2}
\sqrt{2Z^2-2\tilde{\xi}},\\\nonumber
&-Y_{TF}+\vartheta_2^{(D)}=\left[\left(a^2-\frac{f_R'^2 \omega^2
e^{-\lambda}}{4f_R^2
(1-\omega^2)}+\varepsilon_1\right)^{1/2}\right]^\dagger
+\left[a^2-\frac{f_R'^2 \omega^2 e^{-\lambda}}{4f_R^2
(1-\omega^2)}+\varepsilon_1\right]+\\\nonumber
&\frac{1}{2}\left[\sigma+\frac{3\dot{f_R}e^{-\nu/2}(1-\omega^2)^{1/2}}{\omega^2f_R}-\frac{2\omega
f_R' e^{-\lambda/2}}{f_R
(1-\omega^2)^{1/2}}\right]^\star+\frac{1}{3}\left[\left(\theta-\frac{2\dot{f_R}e^{-\nu/2}}{f_R
(1-\omega^2)^{1/2}}-\frac{2f_R'\omega e^{-\lambda/2}}{f_R
(1-\omega^2)^{1/2}}\right)\right.
\\\nonumber&\times\left.\left(\sigma+\frac{3\dot{f_R}e^{-\nu/2}(1-\omega^2)^{1/2}}{\omega^2f_R}-\frac{2\omega
f_R' e^{-\lambda/2}}{f_R
(1-\omega^2)^{1/2}}\right)\right]-\frac{s^1}{r}\sqrt{a^2-\frac{f_R'^2
\omega^2 e^{-\lambda}}{4f_R^2
(1-\omega^2)}+\varepsilon_1}\\\label{40d}
&+\frac{1}{12}\left[\sigma+\frac{3\dot{f_R}e^{-\nu/2}(1-\omega^2)^{1/2}}{\omega^2f_R}-\frac{2\omega
f_R' e^{-\lambda/2}}{f_R (1-\omega^2)^{1/2}}\right],\\\nonumber
&\frac{1}{3f_R}\left(X_T+Y_T-2X_{TF}+Y_{TF}-\eta_1-\eta_2+2\vartheta_1^{(D)}+\vartheta_2^{(D)}\right)\left[\theta-\frac{2\dot{f_R}e^{-\nu/2}}{f_R
(1-\omega^2)^{1/2}}\right.\\\nonumber &\left.-\frac{3f_R'\omega
e^{-\lambda/2}}{f_R
(1-\omega^2)^{1/2}}+\frac{\sigma}{2}+\frac{3\dot{f_R}e^{-\nu/2}(1-\omega^2)^{1/2}}{\omega^2f_R}\right]+\left[\frac{X_T}{2}-X_{TF}
-\frac{\theta_1}{2}+\vartheta_1^{(D)}\right]^\star=-\frac{3\sqrt{2}s^1}{2r}\\\label{41d}
&\times\sqrt{Z^2-\tilde{\xi}},\\\nonumber
&\left[\frac{X_T}{2}-X_{TF}
-\frac{\eta_1}{2}+\vartheta_1^{(D)}\right]^\dagger=\frac{3s^1}{r}\left(X_{TF}-\vartheta_1^{(D)}\right)+\frac{1}{2}
\sqrt{2Z^2-2\tilde{\xi}}\left[\frac{\sigma}{2}+\theta\right.\\\label{42d}
&\left.+\frac{3\dot{f_R}e^{-\nu/2}(1-\omega^2)^{1/2}}{2\omega^2f_R}-\frac{3\omega
f_R' e^{-\lambda/2}}{f_R
(1-\omega^2)^{1/2}}-\frac{2\dot{f_R}e^{-\nu/2}}{f_R
(1-\omega^2)^{1/2}}\right].
\end{align}
The last equation becomes
\begin{equation}\label{43d}
\frac{3m f_R}{r^3}=\frac{X_T}{2}-X_{TF} -\frac{\eta_1}{2}-\frac{\bar{\tau}_0+\bar{\tau}_1}{2}+\vartheta_1^{(D)}.
\end{equation}
Keeping in view Eqs. (97)-(104) for GR in \cite{41}, each of the above equation maintains a correspondence with the said equations with a significant change that our equations relate the kinematical functions with the modified structure scalars (emended using the Palatini's approach).
The Super-Poynting vector and Superenergy can also be restated as
\begin{align}\nonumber
&\bar{W}=\frac{1}{6}\left(X_T^2 +Y_T^2 -N_1\right)+\frac{1}{3}\left(X_{TF}^2+Y_{TF}-N_2\right)+\sqrt{Z^2-\tilde{\xi}},\\\nonumber
&\bar{P}_\beta=\frac{1}{3}\sqrt{2Z^2-2\tilde{\xi}}\left[X_T+Y_T+X_{TF}+Y_{TF}-\eta_1-\eta_2-\vartheta_1^{(D)}-\vartheta_2^{(D)}
\right]s_\beta-\varsigma_\beta.
\end{align}
The values for $N_1$ and $N_2$ are as under
\begin{align}\nonumber
N_1&=\eta^2_1+2\kappa \rho \eta_1+\eta^2_2-2\left(\frac{\kappa \tilde{\rho}}{2}+\frac{3\kappa \hat{P}}{2}\right)\eta_2,\\\nonumber
N_2 &={\vartheta_1^{(D)}}^2+{\vartheta_2^{(D)}}^2+\vartheta^{(D)}_2)+2(E+\frac{\kappa \Pi}{2})\vartheta^{(D)}_2-2\left(\frac{\kappa \Pi}{2}-E\right)\vartheta^{(D)}_2.
\end{align}

\subsection{Active Gravitational Mass in terms of Structure Scalars}

This section deals with restating the expression for active gravitational spherical mass in the context of the structure scalars. Eq. (\ref{27d}) is re-written as
\begin{align}\nonumber
m_T&= (m_T)_\Sigma\left(\frac{r}{r_\Sigma}\right)^3-r^3 \int_r^{r_\Sigma}\frac{\nu' f_R'e^{\frac{\nu-\lambda}{2}}}{2r}dr-r ^3\int_r^{r_\Sigma} \frac{f_R\ddot{\lambda}r^3 e^{\frac{\lambda-\nu}{2}}}{2r}dr
-r^3\int^{r_\Sigma}_r\frac{ e^{\frac{\nu+\lambda}{2}}}{r} \times\\\nonumber &\left(Y_{TF}-\vartheta_2^{(D)}+\frac{\bar{\tau}_1-2\bar{\tau}_2}{\kappa}\right)dr.
\end{align}
For both equilibrium and quasi-equilibrium states, we acquire
\begin{equation}\nonumber
m_T=\int_0^r e^{\frac{\nu+\lambda}{2}}r^2(Y_T-\eta_2)dr.
\end{equation}

\section{Locally Anisotropic Static Spherical Systems}

Sticking to static spherical configurations, we carry out three distinct systematic approaches to arrive at three distinct line elements. Putting into use the stellar equations derived in previous sections, we infer these three different techniques stated below.

\subsection{First Technique}

Putting into use Eqs. (\ref{18d}) and (\ref{43d}), we infer
\begin{equation}\label{44d}
e^{-\lambda}-1=-\frac{2r^2}{3f_R}\left[\frac{X_T}{2}-X_{TF}-\frac{\eta_1}{2}-\frac{\bar{\tau}_0+\bar{\tau}_1}{2}+\vartheta^{(D)}_1\right].
\end{equation}
In the context of static case, Eqs. (\ref{38d}) and (\ref{40d}) deduce the following outcome
\begin{equation}\label{45d}
a=\frac{r}{3s^1}\left(Y_{TF}+Y_T-\eta_2-\vartheta^{(D)}_2\right).
\end{equation}
In terms of metric coefficients and Eq. (\ref{15d}), we infer
\begin{equation}\label{46d}
a=\sqrt{\frac{e^{-\lambda}\nu'^2}{4}-\frac{e^\frac{-\nu-\lambda}{2}f_R'^2}{4f_R^2}}; \quad s^1= e^{-\lambda/2}.
\end{equation}
Substituting Eq. (\ref{46d}) back into (\ref{45d}), we render
\begin{align}\nonumber
e^\nu&=c_1 exp\left[\int\frac{2r}{3}\left[1-\frac{2r^2}{3f_R}\left(\frac{X_T}{2}-X_{TF}-\frac{\eta_1}{2}+\vartheta^{(D)}_1
-\frac{\tilde{\tau}_0+\tilde{\tau}_1}{2}\right)\right]^{-1}\times
\left[\left(Y_T+Y_{TF}\right.\right.\right.\\\nonumber &\left.\left.-\eta_2-\vartheta^{(D)}_2\right)^2+\frac{e^{\frac{-\nu-\lambda}{2}}f_R'^2 }{4f_R^2}\right]^{1/2}dr.
\end{align}
Junction conditions can be utilized to attain the value of the integration constant $c_1$. The line element is yielded as
\begin{align}\nonumber
ds^2&=c_1exp\left[\int\frac{2r}{3}\left[1-\frac{2r^2}{3f_R}\left(\frac{X_T}{2}-X_{TF}-\frac{\eta_1}{2}+\vartheta^{(D)}_1-\frac{\tau_0+\tau_1}{2}\right)
\right]^{-1}\times\left[\left(Y_T+Y_{TF}\right.\right.\right.\\\nonumber &\left.\left.\left.-\eta_2-\vartheta^{(D)}_2\right)^2+\frac{f_R'^2 e^{\frac{-\nu-\lambda}{2}}}{4f_R^2}\right]^{1/2}dr\right]dt^2-\left[1-\frac{2r^2}{3f_R}\left(\frac{X_T}{2}-X_{TF}-\frac{\eta_1}{2}-\right.
\right.\\\nonumber &\left. \left.\frac{\bar{\tau}_0+\bar{\tau}_1}{2}+\vartheta^{(D)}_1\right)\right]^{-1} dr^2-r^2 d\theta^2-r^2 sin^2\theta d\phi^2.
\end{align}
Comparing the above outcome with the line element in Eq. (115) in \cite{41} it is revealed that for static structures in the frame of $f(R)$ gravity by Palatini's approach, the metric also has direct dependence on the effects generated by the existence of the dark source.
\subsection{Second Technique}

Another line element can be procured by using the following strategy. Equations (\ref{20d}), (\ref{31d}) and (\ref{43d}) when combined together, produce the following outcome
\begin{equation}\nonumber
m(r)=r^3\int\left(\frac{X_{TF}}{rf_R}+\breve{\tau}\right)dr+c_2,
\end{equation}
employing the integration of which infers
\begin{equation}\nonumber
m(r)=\frac{r^3}{3f_R}\left(\frac{m'f_R}{r^2}+\frac{\bar{\tau}_0}{2}\right)-\frac{r^3}{3f_R}\left(X_{TF}-\vartheta_1^{(D)}\right).
\end{equation}
From the definition of interior spherical mass,
\begin{equation}\nonumber
e^{-\lambda}=1-2r^2 \left(\int\left[\frac{X_{TF}}{rf_R}+\breve{\tau}\right] dr+ c_2\right),
\end{equation}
where $\breve{\tau}=-\frac{\tau_0}{2r}-\frac{\vartheta_1^{(D)}}{r}$. Again, the junction condition can be used to determine the value for integration constant $c_2$. Putting into use Eq. (\ref{5d}) in static form, we render
\begin{equation}\nonumber
\frac{\kappa P_r}{2}=\frac{f_R}{2}\left[\frac{e^{-\lambda}-1}{r^2}+\frac{\nu' e^{-\lambda}}{r}\right]-\frac{\bar{\tau}_2}{2}.
\end{equation}
Using Eqs. (\ref{18d}), (\ref{33d}) and (\ref{34d}) in combination with (\ref{45d}) and (\ref{46d}), we attain
\begin{equation}\label{47d}
\left(\frac{\kappa P_r}{2}+\frac{mf_R}{r^3}\right)= -\frac{\bar{\tau}_2}{2}+\frac{ e^{-\lambda}\nu' f_R}{2r}=Y_h,
\end{equation}
where
\begin{equation}\nonumber
Y_h=\frac{f_R}{3}\sqrt{\left(Y_T+Y_{TF}-\theta -\vartheta_2^{(D)}\right)^2+\frac{f_R'^2 e^{\frac{-\nu-3\lambda}{2}}}{4r^2 f_R^2}}.
\end{equation}
Integrating Eq. (\ref{47d}), we render
\begin{align}\nonumber
e^\nu=c_3 exp\left[\int\frac{2r Y_h}{f_R}\left(1-2r^2\left[\int\left(\frac{X_{TF}}{rf_R}+\breve{\tau }\right)dr\right]\right)^{-1}dr\right].
\end{align}
The second static line element reads
\begin{align}\nonumber
ds^2&=\left(c_3 exp\left[\int\frac{2r Y_h}{f_R}\left(1-2r^2\left[\int\left(\frac{X_{TF}}{rf_R}+\tilde{\tau }\right)dr+c_2\right]\right)^{-1}dr\right]\right)dt^2\\\nonumber &-\left[1-2r^2 \left(\int\left(\frac{X_{TF}}{rf_R}+\tilde{\tau}\right)dr+c_2\right)\right]^{-1}dr^2-r^2(d\theta^2+sin^2\theta d\phi^2).
\end{align}
Comparing this result with Eq. (123) in \cite{41}, it comes out that the modified structure scalars along with the dark source factors can be used to explicate a static spherical object when working with theories other than GR.
\subsection{Third Technique}

From the reorganization of Weyl scalar $E$, we deduce
\begin{equation}\nonumber
E=-\frac{e^{-\lambda}}{2}\left[\left(\frac{\nu'}{2}\right)^2+\frac{\nu''}{2}+\frac{\lambda'}{2r}+
\frac{1-e^\lambda}{r^2}+
\frac{\nu'}{2}\left(-\frac{\lambda'}{2}-\frac{1}{r}\right)\right].
\end{equation}
Introduction of new variables in the context of metric coefficients provided in the line element (\ref{2d}) infers
\begin{equation}\nonumber
\frac{\nu'}{2}=\frac{u'}{u};\quad\quad\quad y=e^{-\lambda}.
\end{equation}
Under these new variables, a differential equation is ascertained as
\begin{equation}\nonumber
y'+2\left(\frac{u''-\frac{u'}{r}+\frac{u}{r^2}}{u'-\frac{u}{r}}\right)y=\frac{2(1-2r^2E)u}
{\left(u'-\frac{u}{r}\right)r^2}.
\end{equation}
Employing integration w.r.t. $r$ yields
\begin{equation}\nonumber
y=e^{-\int \tilde{k}(r)dr}\left(\int e^{-\int \tilde{k}(r)dr}\tilde{f}(r)dr+c_4\right).
\end{equation}
The functions $\tilde{k}(r)$ and $\tilde{f}(r)$ are given as
\begin{equation}\nonumber
\tilde{k}(r)=2 \frac{d}{dr}\left[ln\left(u'-\frac{u}{r}\right)\right],
\quad\quad \tilde{f}(r)=\frac{2(1-2r^2E)u} {\left(u'-\frac{u}{r}\right)r^2}.
\end{equation}
Switching to old variables, we obtain a first order differential equation in $\nu$ as
\begin{equation}\label{48d}
-\frac{1}{r}+\frac{\nu'}{2}=\frac{e^{\lambda/2}}{r}\sqrt{(1-2Er^2)+c_4r^2e^{-\nu}+
r^2e^{-\nu}\int\frac{e^\nu}{r^2}(2r^2E)'dr}.
\end{equation}
Introducing another new variable $z$ in terms of $\nu$ as
\begin{equation}\label{49d}
e^\nu=\frac{e^{2\int zdr}}{r^2},
\end{equation}
which deduces
\begin{equation}\label{50d}
z(r)=\frac{1}{r}+\frac{\nu'}{2}.
\end{equation}
Using Eqs. (\ref{49d}) and (\ref{50d}) into (\ref{48d}), we render
\begin{equation}\label{51d}
z(r)-\frac{2}{r}=\frac{e^{\lambda/2}}{r}\sqrt{(1-2Er^2)+c_4r^4e^{-\int
2z(r)dr}+ r^4e^{-\int 2z(r)dr}\int\frac{e^{-\int
2z(r)dr}}{r^4}(2r^2E)'dr}.
\end{equation}
Eqs. (\ref{3d}) yield
\begin{align}\nonumber
\kappa\Pi&=f_R\left[e^{-\lambda}\left(-\frac{\nu''}{2}-\left(\frac{\nu'}{2}\right)^2+
\frac{\nu'}{2r}+\frac{1}{r^2}\right)+\frac{\lambda'e^{-\lambda}}{2}\left(\frac{\nu'}{2}+
\frac{1}{r}\right)-\frac{1}{r^2}\right]+(\bar{\tau}_2-\bar{\tau}_1).
\end{align}
Placing the values for $y$ and $z$ above, we deduce
\begin{align}\label{52d}
y'+y\left[\frac{2z'}{z}+2z-\frac{6}{r}+\frac{4}{r^2z}\right]=-\frac{2}{z}\left[\frac{\kappa\Pi}{f_R}
+\frac{1}{r^2}-\frac{(\bar{\tau}_2-\bar{\tau}_1)}{f_R}\right].
\end{align}
For the metric coefficient $\lambda$, we attain
\begin{align}\label{53d}
e^{\lambda(r)}=\frac{z^2
e^{\int\left(2z+\frac{4}{zr^2}\right)dr}}{r^6\left[-2\int\frac{z}{r^8}\left(\frac{\kappa\Pi
r^2}{f_R}+1-\frac{(\bar{\tau}_2-\bar{\tau}_1)r^2}{f_R}\right)e^{\int\left(2z+\frac{4}{zr^2}\right)dr}dr+c_5\right]}.
\end{align}
The scalar variables $X_{TF}$ and $Y_{TF}$ are utilized and  Eqs. (\ref{51d}) and (\ref{53d}) are restated as
\begin{align}\nonumber
z(r)&=\frac{2}{r}+\frac{e^{\lambda/2}}{r}\left[1-r^2\left(Y_{TF}-X_{TF}-\vartheta^{(D)}_1-\vartheta^{(D)}_2\right)+
r^4e^{-\int2zdr}\left(c_4+\int\frac{e^{\int2zdr}}{r^4}\times\right.\right.\\\label{54d} &\left.\left.\left(r^2(Y_{TF}-X_{TF}-\vartheta^{(D)}_1-\vartheta^{(D)}_2)\right)
'dr\right)\right]^{1/2},\\\label{55d}
e^{\lambda(r)}&=\frac{z^2e^{\int\left(2z+\frac{4}{zr^2}\right)dr}}{r^6\left[-2\int\frac{z}{r^8}\left(\frac{r^2}{f_R}(Y_{TF}
+X_{TF}+\vartheta^{(D)}_1+\vartheta^{(D)}_2)+1-\frac{(\bar{\tau}_2-\bar{\tau}_1)r^2}{f_R}\right)e^{\int\left(2z+\frac{4}{zr^2}\right)dr}dr+c_5\right]}.
\end{align}
The value for the variable $z$ can be ascertained by using Eq. (\ref{55d}) into (\ref{54d}) from which the metric coefficient $\nu$ is easily determinable. Also, utilizing the value for $z$ in Eq. (\ref{55d}), we can determine the metric coefficient $\lambda$. The last outcome develops correspondence with Eq. (136) of \cite{41} with a difference that in our case, the metric coefficient $e^{\lambda(r)}$ can be fully described by the modified structure scalars and the dark source factors.

\section{Outline of the Results}

From the recently conducted astrophysical research on Cosmic
Microwave Background, Supernovae type Ia and numerous large scale
configurations, it has been revealed that our Universe is enduring
an accelerated expansion phase and its explication demands the
inclusion of Dark Energy in gravity theories. In this regard, mostly
$f(R)$ theories are exploited because they open up a new way to
contemplate mysterious forces in nature by putting into
use a generic function of curvature invariant, i.e., Ricci scalar in
the Lagrangian. Variation of such a kind of Lagrangian along with
matter action renders second order gravitational equations that are
free from any singularity in addition with few supplementary terms
arising due to $f(R)$ emendations. Palatini $f(R)$ theory is supreme
when compared to other approaches because is meets all the criteria
of Solar system tests and supplies accurate Newtonian limit.

Summing up the findings of the article, we can say that we have
inquired the contribution of structure scalars in the assessment of
composition and time-evolution of a spherically symmetric star. To
achieve this objective, we devise the devise the modified
gravitational field equations by putting into use the Palatini's
technique of variating the Einstein-Hilbert action. We pay heed to a
non-comoving reference frame because it has a substantial role in
explicating the mysterious secrets of evolution of our Universe. The
relativistic fluid content under consideration is assumed to stress
anisotropy attribute with emission of heat radiations. We carry out
the evaluation of various fluid kinematical variables and devised
numerous equations that assist in the inspection of structure
composition and its evolution as the time passes.

By orthogonally decomposing the Riemann tensor, we developed three
tensorial quantities which in turn generate five modified scalar
functions $X_T,~Y_T,~X_{TF},~Y_{TF}$ and $Z$. The scalar $X_T$
deduced from te trace component of $X_{\varrho\beta}$ is associated
with the description of energy density with some supplementary terms
emerging due to Palatini $f(R)$ emendations. Its trace-free
component $X_{TF}$ administers the energ density inhomogeneity as
noticed from Eq. (\ref{42d}) (When $Z=0$, we acquire a differential
equation in $\tilde{\rho}$ which by making use of the regular center
condition implies that the change in energy density is zero if and
only if $X_{TF}$ is zero). From the trace and trace-free components
of $Y_{\varrho\beta}$ i.e., $Y_T$ and $Y_{TF}$ it is inferred that
$Y_{TF}$ explicates the impact of anisotropic stresses as well as
inhomogeneous energy density on the active gravitational mass of the
sphere. $Y_T$ appears in the expression for $m_T$ so it can be
described as the active gravitational mass density. The scalar $Z$
is connected with the heat dissipative flux of the fluid source. All
these five scalars facilitate in demonstrating the singularity
formation in the astronomical objects like Blackholes. Taking
account of spherical static symmetry, we developed three distinct
line elements stemming from the general one (provided in Eq.
(\ref{2d})) to elucidate static spherical stars by employing
structure scalar functions.

\end{document}